\pdfoutput=1

\documentclass[onecolumn,12pt,superscriptaddress]{revtex4}
\usepackage{amsfonts,amssymb,amsmath,mathbbol}              
\usepackage{epstopdf}
\usepackage{mathbbol}              
\usepackage{graphics,graphicx,epsfig,ulem} 
\usepackage{eurosym}
\graphicspath{{my_diagrams/}}
\usepackage{color}
\usepackage{gensymb}
\usepackage{subfigure}
\usepackage{graphicx}
\usepackage{natbib}
\usepackage{bbm}
\usepackage{textcomp}

\begin{document}

\textheight=23.75cm

\title{Effect of gain and phase errors on SKA1-low imaging quality from 50\--600\,MHz} 

\author{D. R. Sinclair}

\email{David.Sinclair@astro.ox.ac.uk}

\affiliation{Department of Physics, University of Oxford, Keble Rd, Oxford, UK, OX1 3RH}

\author{F. Dulwich}

\affiliation{Oxford e-Research Centre, University of Oxford, Keble Rd, Oxford, UK, OX1 3QG}

\author{B. Mort}

\affiliation{Oxford e-Research Centre, University of Oxford, Keble Rd, Oxford, UK, OX1 3QG}

\author{M. E. Jones}

\affiliation{Department of Physics, University of Oxford, Keble Rd, Oxford, UK OX1 3RH}

\author{K. Grainge}

\affiliation{School of Physics and Astronomy, University of Manchester, Oxford Rd, Manchester, UK, M13 9PL}

\author{E. de Lera Acedo}

\affiliation{Cavendish Laboratory, University of Cambridge, JJ Thomson Avenue, Cambridge, UK, CB3 0HE}

\begin{abstract}
\vspace{1em}
Simulations of SKA1-low were performed to estimate the noise level in images produced by the telescope over a frequency range 50\--600\,MHz, which extends the 50\--350\,MHz range of the current baseline design. The root-mean-square (RMS) deviation between images produced by an ideal, error-free SKA1-low and those produced by SKA1-low with varying levels of uncorrelated gain and phase errors was simulated. The residual in-field and sidelobe noise levels were assessed. It was found that the RMS deviations decreased as the frequency increased. The  residual sidelobe noise decreased by a factor of ${\sim}\,5$ from 50 to 100\,MHz, and continued to decrease at higher frequencies, attributable to wider strong sidelobes and brighter sources at lower frequencies. The thermal noise limit is found to range between ${\sim}\,10 - 0.3$  \textmu Jy and is reached after ${\sim}\,100-100\,000$ hrs integration, depending on observation frequency, with the shortest integration time required at ${\sim}\,100$\,MHz.

\end{abstract}

\maketitle

\section{Introduction}
The SKA low frequency aperture array, SKA1-low, is proposed to cover the frequency range 50\--350\,MHz. At the lower end of the frequency band, images are created with relatively wide station beams, and hence also wide near-in sidelobes. This could present difficulties in image processing to create the large imaging dynamic range desired by the SKA. 

At the upper end of the SKA-low frequency band, it has been suggested that the range could be extended to 650\,MHz by sampling in the second Nyquist zone of detected signals, which would be a relatively inexpensive upgrade to the current instrument design \cite{LFAA_techdesc}. However, there are concerns over the fidelity of the images that would be produced at such high frequencies due to the very sparse sampling of the aperture array. The original specification of the aperture array called for a critical sampling frequency of 110 MHz  \cite{Faulkner2013}, which is an appropriate concept for a regular array. The currently proposed array has irregular spacings with an average spacing of 1.93\,m and an approximate dense-sparse transition frequency of 77\,MHz.

The aim of this paper is to investigate how imaging is affected across a frequency range of 50\--600\,MHz by simulating the effect of phase and gain errors on the images produced by an interferometer matching the SKA1-low specifications. This investigation builds on the work of Razavi-Ghods et al. \cite{Razavi2014} which assesses the far-sidelobe confusion noise of sources in an ideal SKA1-low telescope. In addition to modelling phase and gain errors, baselines beyond the SKA1-low core are included and the frequency range extended down to 50\,MHz.

Phase and gain errors can be introduced to the signal processing chain from a range of sources. These can include errors in the antenna positions, variations in the precision of mass-produced electronic components and quantisation errors. Discussions of the effects of errors on metrics such as the sidelobe levels, directivity and beam pointing accuracy of an aperture array can be found in the literature (for example, \cite{Mailloux}, \cite{carver1973}, \cite{Schediwy2010}).

\section{Methods}
\label{sec:methods}
The Nyquist sampling theorem states that a band-limited function can be completely determined by sampling it at a frequency of twice the bandwidth. In spatial terms this corresponds to sampling at every half-wavelength of the shortest wavelength present. This means that the distance between antennas in an aperture array is significant: if antennas are spaced further than half the observed wavelength apart, the station beam is spatially aliased, resulting in strong sidelobes known as grating lobes (see, for example, \cite{Hansen}). The frequency at which this transition occurs is the critical sampling frequency.

A source located in a sidelobe of a radio telescope will be detected at a level dependent on the sidelobe strength. It is theoretically possible to remove the signal from sources in far-out sidelobes in synthesised images with post-processing. However, this requires a perfect understanding of the telescope's behaviour, and of the sources in the sky. When uncertainties are present in the signal processing chain, the resultant data are perturbed from the error-free case. By investigating the differences between images from an ideal (error-free) telescope and those from a telescope with errors, the effect of errors on the images can be quantified.

The OSKAR simulator \cite{OSKAR2014} provides a flexible environment in which large-scale simulations can be conducted, by utilising the Radio Interferometer Measurement Equation \cite{Hamaker1996, Smirnov2011}. OSKAR was used to generate simulated visibilities for all cases. OSKAR is the principal tool used in this paper to simulate visibilities, while CASA was used to make images of the simulated data. 

\begin{figure*}[htp]
\subfigure [~Layout of the telescope model used. There are 1024 stations, with 50\% within 600\,m of the centre and 75\% within 1\,km. There are three spiral arms of 15 stations each.]{\label{fig:tel_layout}\includegraphics[width=0.45\linewidth]{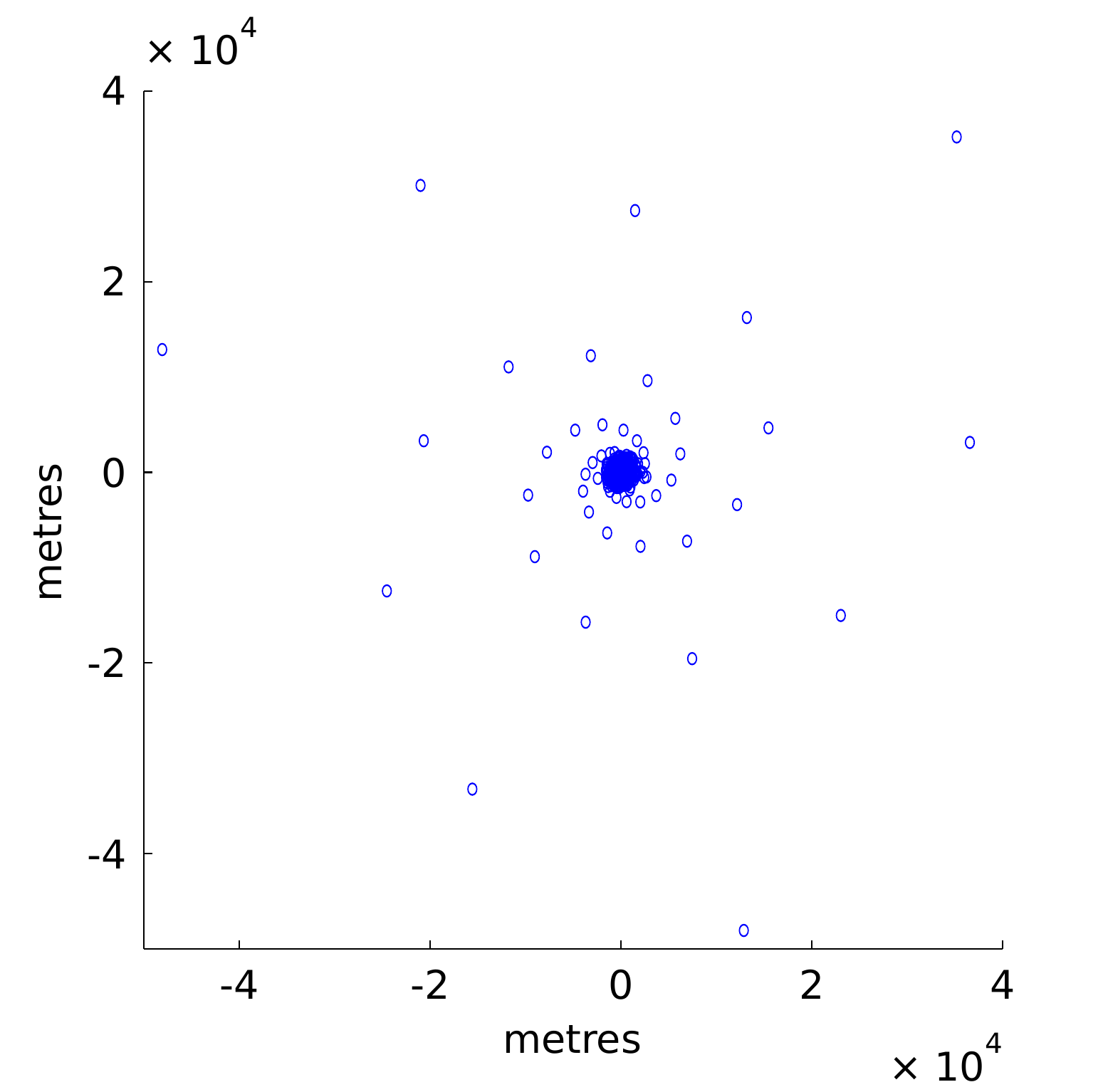}}
\subfigure [~Layout of one station. 768 ($256\times3$) elements within a $\sqrt{3} \times 35$\,m diameter circle, randomly distributed, but separated by a minimum of 1.35\,m (critically sampled at 110\,MHz). Every other station has the same layout, but rotated around the centre by a unique angle.]{\label{fig:stat_layout}\includegraphics[width=0.45\linewidth]{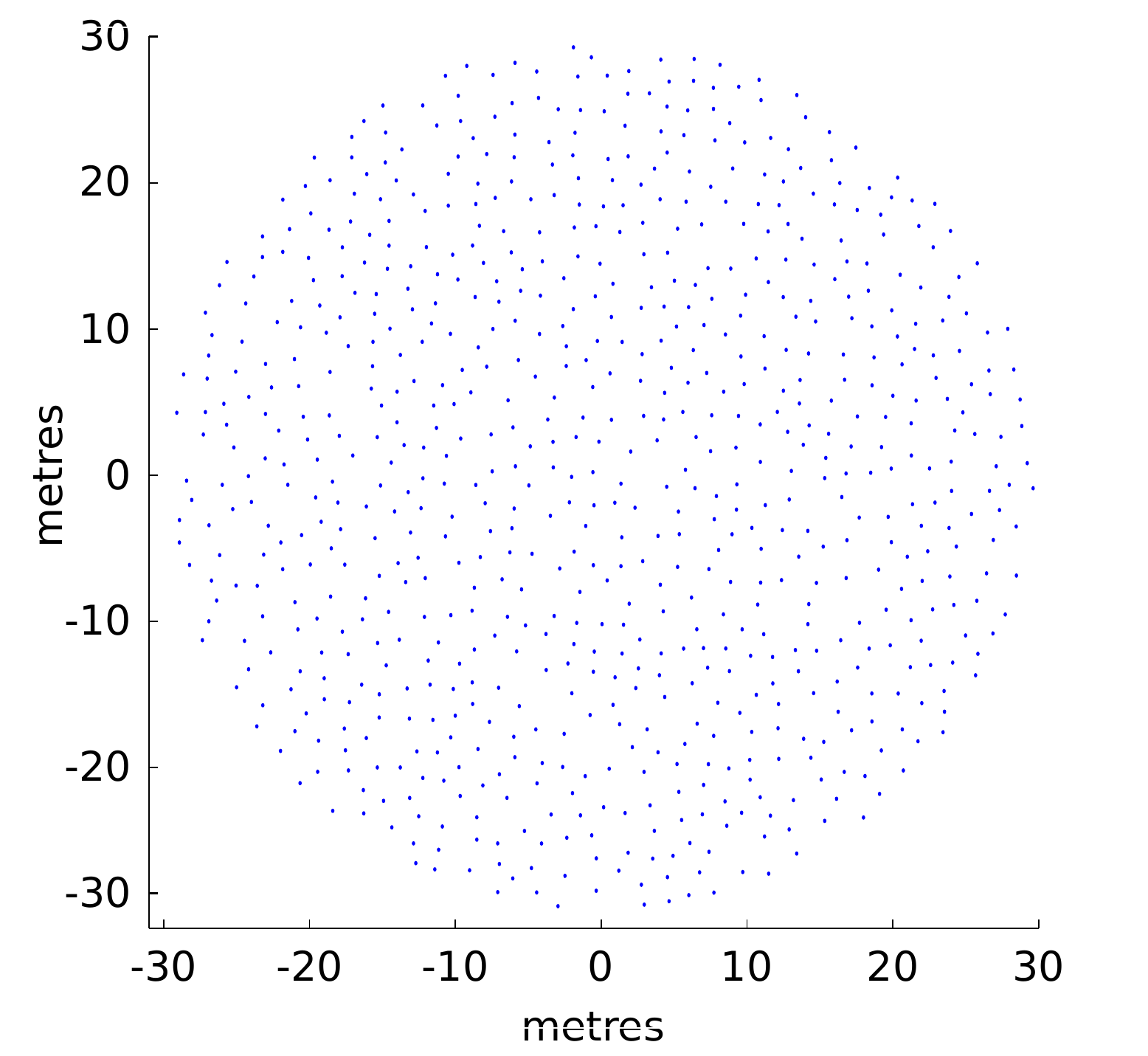}}

\subfigure [~Snapshot \textit{uv} coverage (at the zenith) of the telescope layout.]{\label{fig:uv_cov}\includegraphics[width=0.45\linewidth]{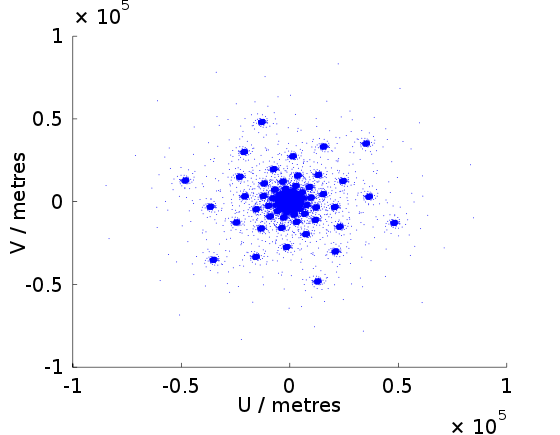}}
  \caption{The telescope model used for these simulations. Station locations conform to the stipulations in the requirements specification \cite{level1reqspec}.}
  \label{tel_layouts}
\end{figure*}

To simulate SKA1-low, a telescope model was created with station locations matching the description given in the original requirements specification \cite{level1reqspec}. The layout and instantaneous \textit{uv} coverage are shown in Fig \ref{tel_layouts}. 

The requirements specification describes a telescope containing 1024 35\,m diameter stations, with 256 antennas per station. Here 768 ($3 \times 256$) antennas per station were used, giving stations a factor $\sqrt{3}$ times the diameter specified. The station beams were apodised using a Hann window such that their half power beam width (HPBW) was equal to a 256-antenna unapodised station. This is discussed in \cite{Grainge2014} but we will briefly summarise it here.

The baseline design of 35\,m diameter separate stations assumes all antennas within the station have equal weight, in order to achieve the required total effective area of the array. However, a station with uniform weighting generates high sidelobes in the station beam, which significantly increases signal power from sources in the sidelobes. Apodising small stations to improve the sidelobe levels results in reduced overall collecting area and larger station beams than desired, where as a scheme of apodised overlapping stations in a contiguous distribution of antennas can give efficient use of collecting area along with good beam performance. The scheme used in this memo gives a good approximation to this concept in so far as utilising apodised stations to reduce sidelobe levels, but a contiguous distribution of antennas in not used. A more accurate version of this scheme will be described in a future memo and in an Engineering Change Procedure (ECP) document.  

A random antenna layout was used which is common to each station, but each of the 1024 stations were rotated by an  angle, $(360n / 1024)^{\circ}$, where $n$ is a unique integer between 0 and 1023. In line with the requirements specification, log-periodic antenna elements with a directivity of almost 8 dBi, were used (a 50\,MHz antenna model was not available, so the element response at 100\,MHz was used; log-periodic antennas are designed to have approximately the same antenna beam pattern across their frequency range). Antenna elements had a minimum separation of 1.35\,m, corresponding to a critical sampling frequency of ${\sim}\,110$\,MHz specified in the original baseline design. 

We define two metrics for measuring the impact of uncorrelated gain and phase errors on imaging: residual in-field noise (RIN) and residual sidelobe noise (RSN). These are RMS deviations between a dirty image made from an ideal, error-free observation and an image from an identical observation with gain and/or phase errors in the signal processing chain. RIN occurs from sources located within the HPBW of the station beam and RSN occurs from sources in the station beam's sidelobes. 

\begin{figure*}
    \centering
    \includegraphics[width=0.8\textwidth]{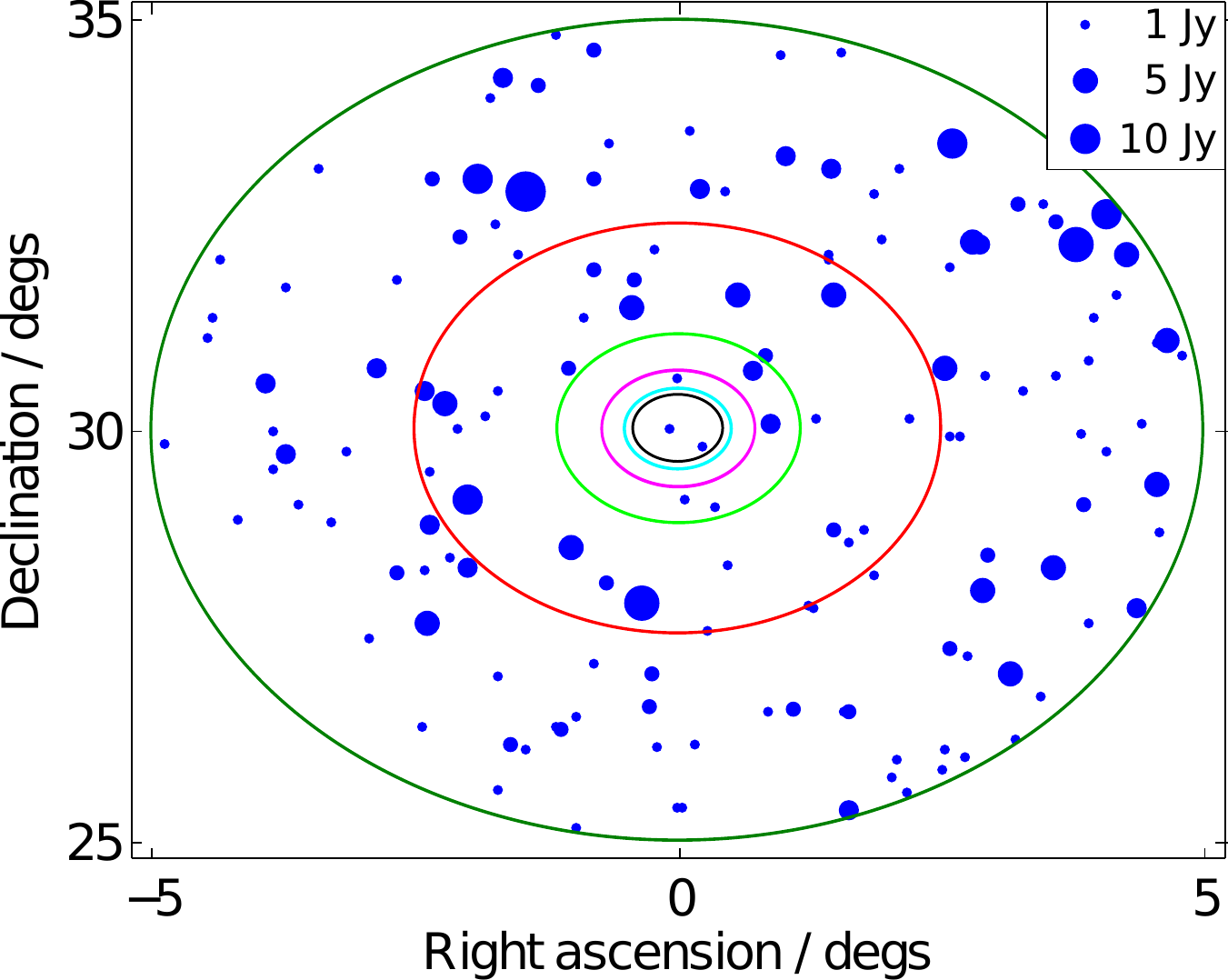}
    \caption{Field of view observed by the station beam of the telescope. The dots represent sources in the VLSS sky model, with their size corresponding to the source strength, as shown in the legend. Source fluxes shown at 74\,MHz; there is a power law spectral index of -0.7 applied to higher frequencies. The outer (dark green) circle shows the field of view at 50\,MHz, with subsequent internal circles increasing in frequency (100, 220, 350, 500, 600\,MHz). At each frequency, sources beyond the field of view were removed from the sky model when investigating the effect of phase and gain errors on sources in the station beam.}
    \label{fig:VLSS_in_beam_FoV}
\end{figure*}

Four sky models were prepared for use in these simulations, which were divided into two types. The first type of sky model had sources placed only in the main lobe of the station beam: no sources were placed more than the HPBW away from the observation's phase centre, allowing the RIN to be measured. This boundary distance was decreased proportionally with the reduction in field of view at higher frequencies. The second type of sky model did the opposite; all sources within twice the HPBW of the station were removed, but sources populated the rest of the sky, allowing RSN to be investigated. Further details of the sky models are given as follows:

\begin{itemize}
	\item{RIN: sources only within the HPBW}
	\begin{itemize}
		\item{Model 1: A grid of nine 1\,Jy sources. The on-sky separation between the sources was scaled with the observation frequency, such that the nine sources were located on the same image pixels at every observation frequency.}
		\item{Model 2: A point-source sky model using data from the VLA Low-Frequency Sky Survey (VLSS). All sources outside the station station beam were removed. The VLSS was measured at 74\,MHz; a spectral index of  $\alpha = -0.7$  (with $S(\nu) \propto S^{\alpha}$, where $S$ is the flux) was applied to all sources to determine their fluxes at different frequencies. This spectral index is within the typical range for synchrotron radiation in this frequency range (see, for example, \cite{BurkeSmith}). Fig \ref{fig:VLSS_in_beam_FoV} shows how the sources included in the sky model changed with frequency.}
	\end{itemize}
	\vspace{1em}
	\item{RSN: No sources within twice the HPBW.}
	\begin{itemize}
		\item{Model 3: A grid of 1\,Jy sources placed at 10$^{\circ}$ intervals in RA, Dec across the sky, except in the direction of the synthesised beam. Source separation did not scale with frequency.}
		\item{Model 4: The VLSS sky model, as above, but with all sources within the HPBW removed.}		\end{itemize}
\end{itemize}

The VLSS sky model had the six brightest sources removed (Cassiopeia A, Cygnus A, Taurus A, Hercules A, Virgo A, Hydra A), which account for approximately 40.5\,kJy of total flux density at 74\,MHz. These six strong radio sources had the potential to unduly distort the results if they remained. For example, if due to the combination of telescope location, observation direction and observation time, Cassiopeia A were located in a grating lobe, it could significantly affect the measured results compared to a weaker source, which are far more common and thus likely to be in the grating lobe. All bright sources, however, will need to be well understood and accounted for when conducting actual observations with the SKA. 

Simulations used a phase centre at the zenith (except where stated otherwise) and the telescope centred at (0,\,30$^{\circ}$) in (longitude, latitude) the approximate latitude of the SKA1-low site (albeit in the opposite hemisphere; this was to allow the use of the northern sky VLSS catalogue as a sky model). Simulations were run at 50, 100, 220, 350, 500 and 600\,MHz, each of 13.7\,s integration (except where otherwise stated) and 100\,kHz bandwidth. 13.7\,s corresponds to approximately 1 milliradian of earth rotation.  Images were made of Stokes I.

Uncorrelated uncertainties in the gain and phase were added to the signals ÔdetectedÕ at each antenna. These represent the cumulative errors acquired along the signal chain. The uncertainties were assigned from random values selected from Gaussian distributions and were different for each antenna. The values were kept constant throughout the integration time, whether it was a 13.7\,s snapshot or ${\sim}\,1$\,hr long observation. However, the error values were regenerated for repeated iterations to determine errors using the Monte Carlo method. The standard deviations of the uncertainties are the values given in this paper; for instance a phase error of 1$^{\circ}$ means that the phase value applied to an antenna is $\phi \pm \phi_{\textrm{err}}$, where $\phi$ is the ideal phase value and $\phi_{\textrm{err}}$ is the error value, which is randomly selected from a Gaussian distribution of standard deviation 1$^{\circ}$. Likewise a gain error of 10\% represents an additive error selected from a Gaussian distribution of 0.1 times the true value. For large gain errors this means that the gain is potentially negative. However, this is only significant for gain errors of order unity, which are unrealistic, but are included to show the upper bound on the effect of such errors. 

It is useful to compare the RMS deviations to the thermal noise level of the observations. The thermal noise level is the weak-source limit of the telescope, per polarization, and is given by

\begin{equation} \label{equation:thermalnoise}
S_{\textrm{min}} = \frac{1}{\eta_{s}}\frac{SEFD}{\sqrt{2\,\Delta\nu\,t}}\,,
\end{equation}
where $\eta_{s}$ is the system efficiency factor, $\Delta \nu$ is the bandwidth and $t$ is the integration time. $SEFD$ is the system equivalent flux density and is calculated by

 \begin{equation} \label{equation:SEFD}
SEFD = T_{sys}\frac{2k_{B}}{\eta_{A}A_{e}}\,,
\end{equation}
where $k_{B}$ is the Boltzmann constant, $\eta_{A}$ is the antenna efficiency (assumed to be 0.9) $A_{e}$ is the effective area of the antenna and $T_{sys}$ is the system temperature \cite{TaylorCarilliPerley}. 

The SKA1 baseline design \cite{newbaselinedesign} gives a model for $A_{e}$ per antenna of $3.2\,\textrm{m}^{2}$ below 110\,MHz and proportional to $\lambda^{2}$ above. This model has been followed in this paper.

The baseline design also gives a model for $T_{sys}$:
\begin{equation} \label{equation:Tsys}
T_{sys} = T_{sky} + T_{rcvr}\,,
\end{equation}
 where $T_{sky}$ is the sky temperature and $T_{rcvr}$ is the receiver temperature and is modelled as $0.1 T_{sky} + 40\,\textrm{K}$ in the baseline design.

The Milky Way is the brightest radio background at these frequencies and thus its temperature can be considered as a good approximation for $T_{sky}$. The Haslam 408\,MHz all sky map  \cite{Haslam1,Haslam2} provides temperatures for the galactic foreground at this frequency. By measuring the average value of pixels within a station beam-sized area, $T_{sky}$ can be determined for any telescope pointing direction in the sky. 

The average temperature within a station beam-sized area that was colder than average, along with the temperature in an average temperature area of sky were determined   

Temperatures at other frequencies were calculated by $T(\nu _{2}) = T(\nu _{1}) \times (\nu_{2} / \nu_{1})  ^{\alpha}$ for $\alpha =$ -2.4, -2.55 and -2.7, giving a range of temperature values. 

By combining the range of temperatures given by the average and cold patches of sky, along with the different values of $\alpha$, upper and lower bounds on $T_{sky}$ were determined.

The $A_{e}/T_{sys}$ found above are consistent with the range of values given in  E. de Lera Acedo \textit{et al.} \cite{Acedo2014}, which discusses improvements to the models for $A_{e}/T_{sys}$. 

The thermal noise level is the theoretical best weak-source sensitivity of the telescope. Hence if the noise from gain and phase errors exceeds this level, the performance of the telescope will be impaired. A comparison of the two provides an upper bound on the allowable gain and phase errors.

Due to computational resource limits, simulations were mostly run with snapshot observations of 13.7\,s duration. The RMS thermal noise level in a radio astronomy observation will decrease with the square-root of the length of observation. One may also expect the RSN to also decrease with $t^{0.5}$, at least initially, as the rotation of the earth moves the many sources through the sidelobe errors of the telescope in a quasi-random way. This effect should continue for around twelve hours until the \textit{uv} tracks of the baselines begin to repeat themselves and the new noise components are not independent of those previously observed. To check that this occurred, further observations of the RSN using the VLSS sky model were simulated with integration times of up to an hour.    

These simulations are modelling only the effect of gain and phase errors and not all the causes of noise. Preliminary simulations of the RMS noise when accounting for different effects suggest the noise may decrease at a slower rate than $t^{0.5}$ \cite{BandFpersonal}. 

\section{Results and Discussion}
\subsection{Residual in-field noise}
\begin{figure*}
\subfigure [~Station beam pattern of ideal telescope.]{\label{fig:ideal_station_beampattern}\includegraphics[width=0.49\linewidth]{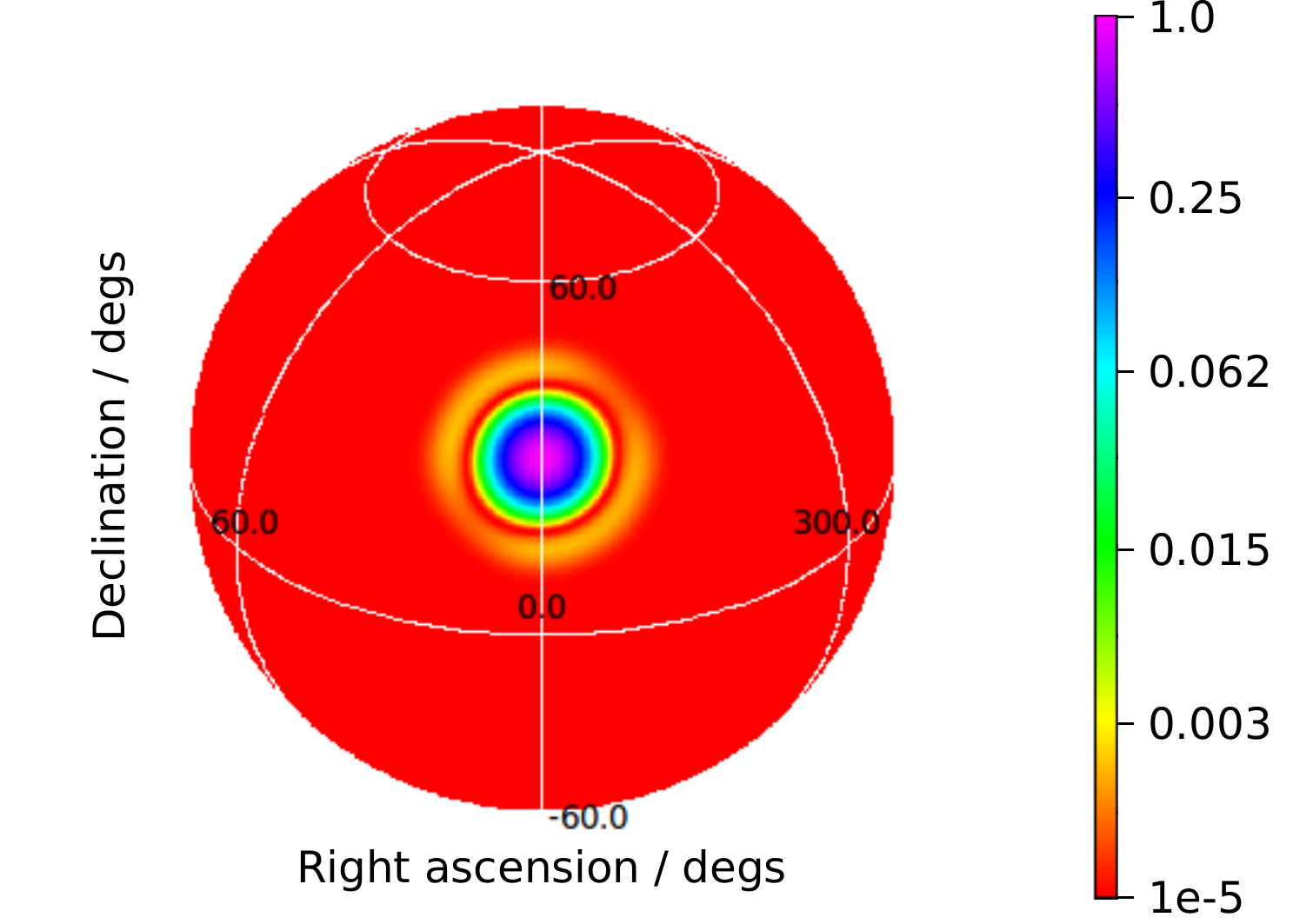}}
\subfigure [~Station beam pattern of telescope\newline with gain errors of 1.]{\label{fig:error_station_beampattern}\includegraphics[width=0.49\linewidth]{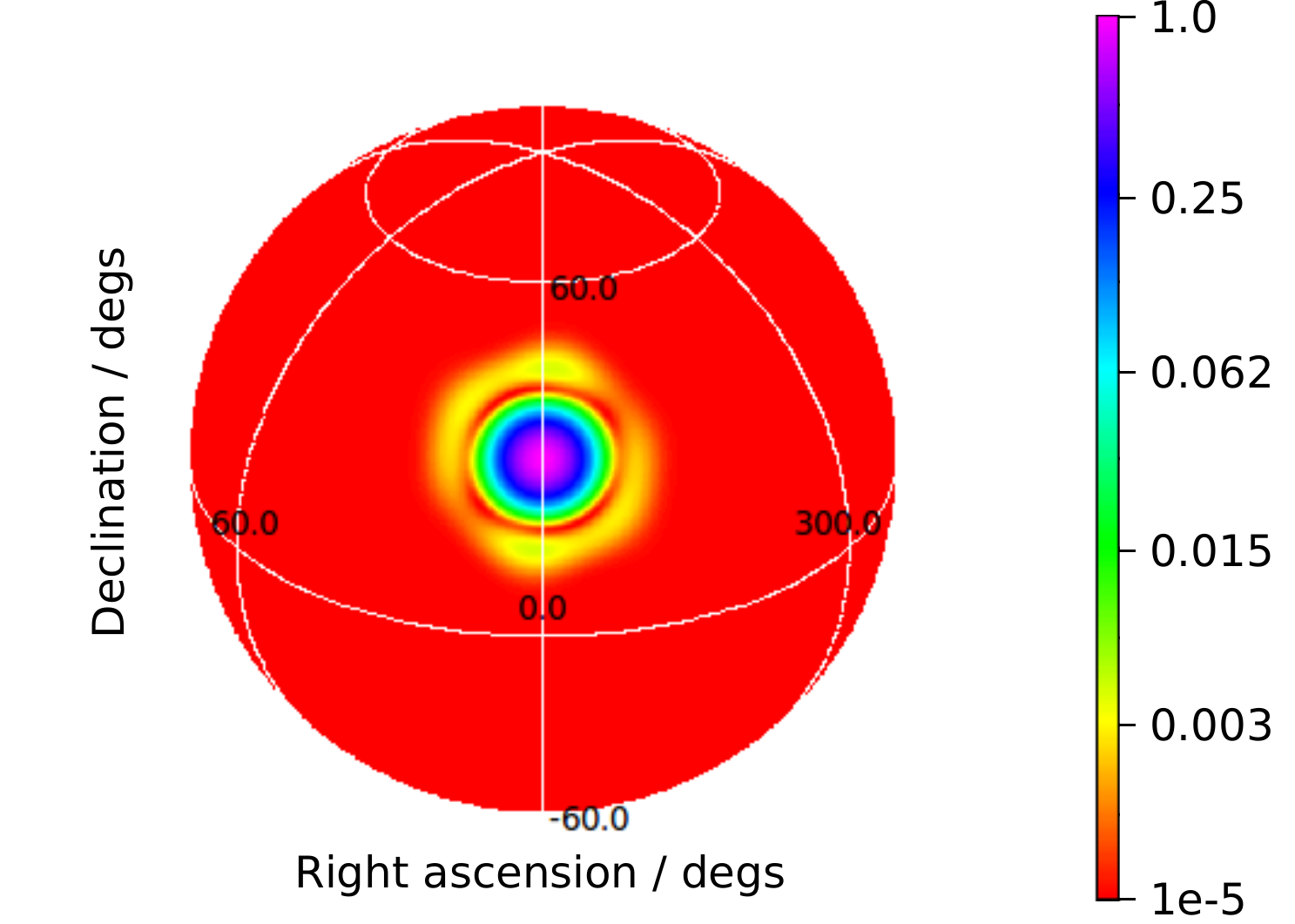}}

\vspace{1em}

\subfigure [~Difference between \subref{fig:ideal_station_beampattern} and \subref{fig:error_station_beampattern}.]{\label{fig:diff_station_beampatterns}\includegraphics[width=0.49\linewidth]{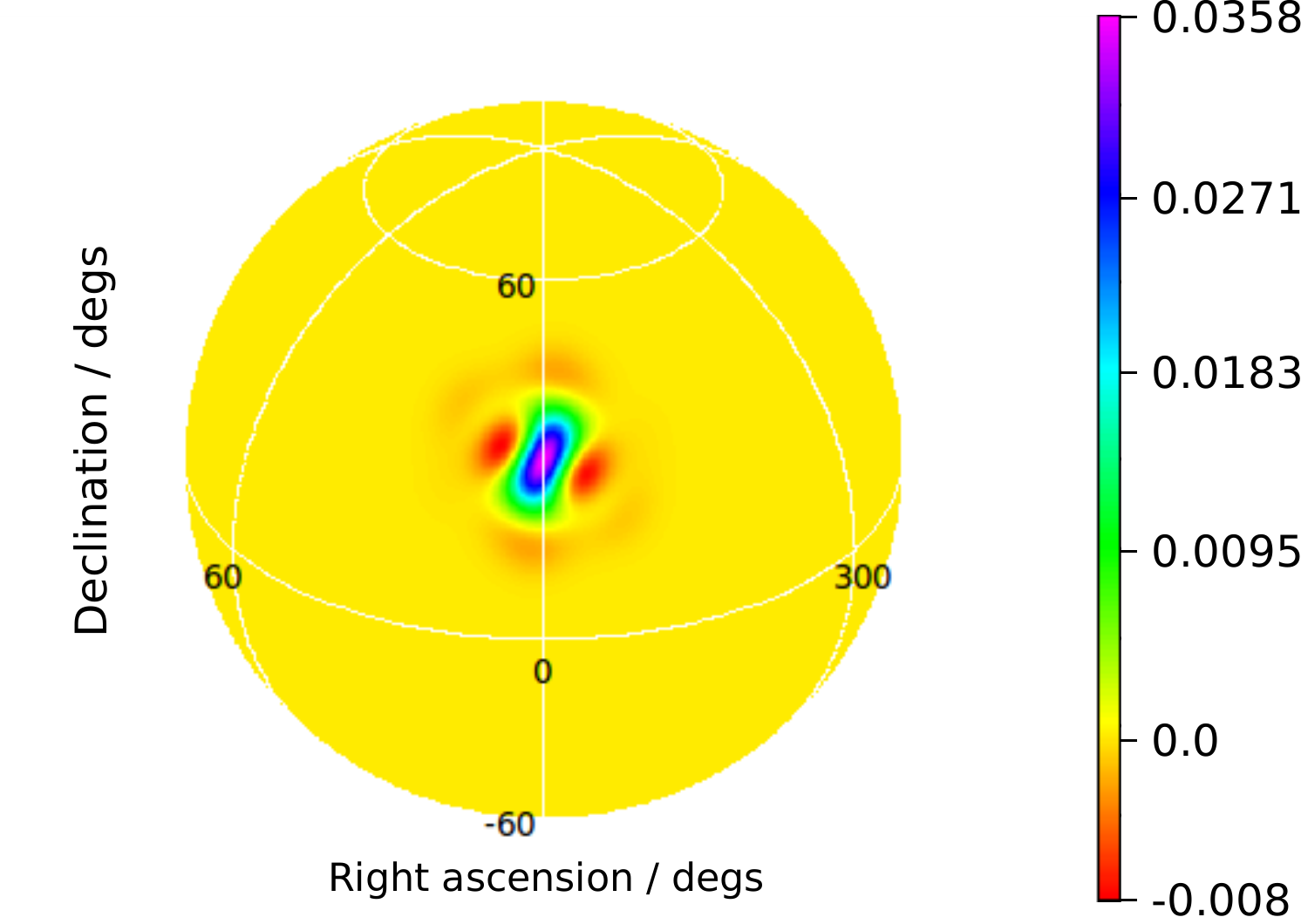}}
  \caption{Station beam patterns for ideal telescope, one with errors, and the difference between the two. The colour scale shows the gain of the beam. Note the change in scales between \subref{fig:ideal_station_beampattern} and \subref{fig:error_station_beampattern} (logarithmic), and \ref{fig:diff_station_beampatterns} (linear).}
  \label{fig:station_beampatterns}
\end{figure*}

The addition of uncorrelated gain and phase errors to the antennas results in distortions of the station beam patterns, which served to add errors to the images, ultimately increasing the RMS deviation between synthesised images from a telescope with errors and images from one without. An example of an ideal station beam pattern, a beam pattern with gain errors, and the difference between the two, are shown in Fig \ref{fig:station_beampatterns}. 

\begin{figure*}
\subfigure []{\label{fig:dense_amperr}\includegraphics[width=0.46\linewidth]{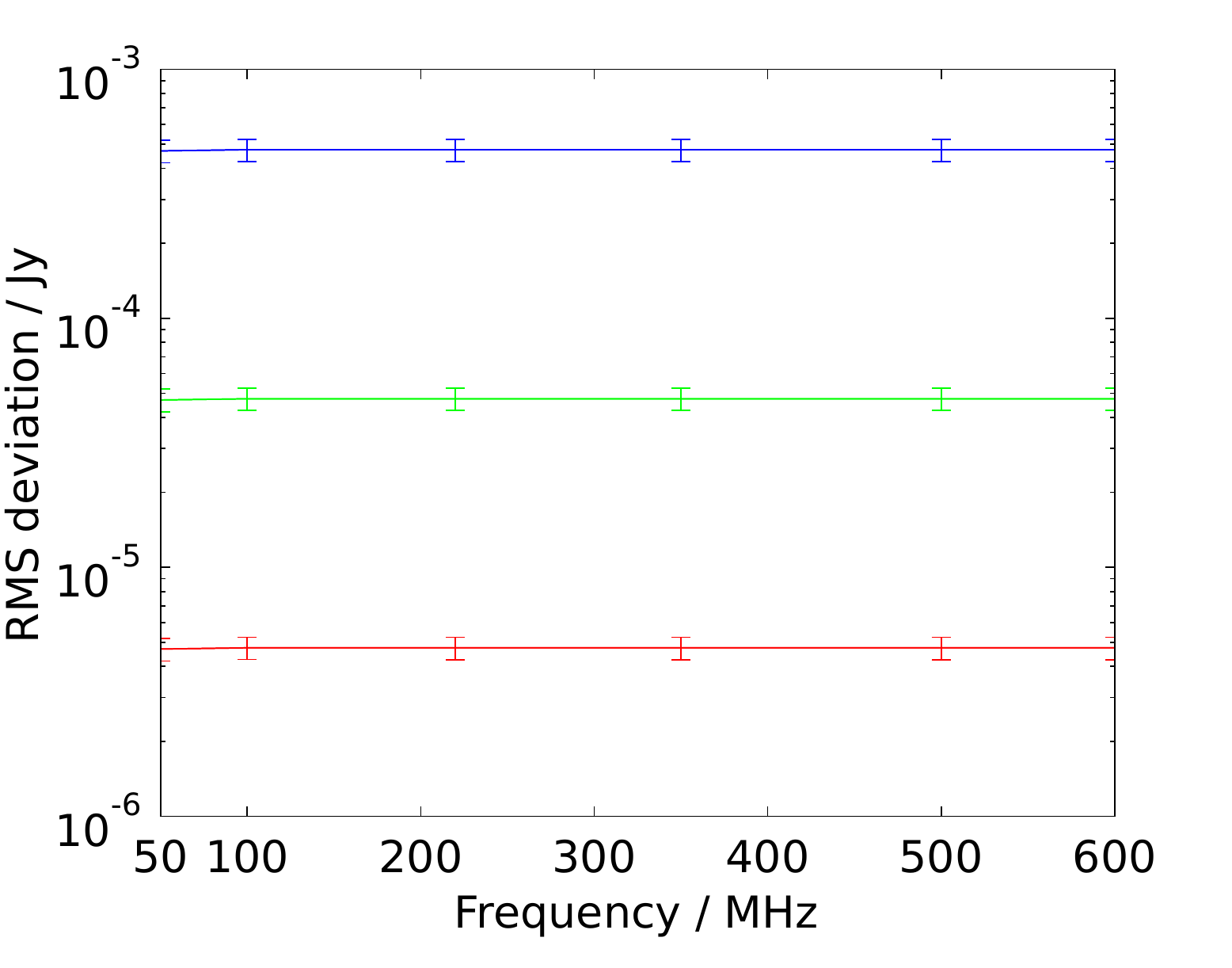}}
\subfigure []{\label{fig:dense_phaseerr}\includegraphics[width=0.46\linewidth]{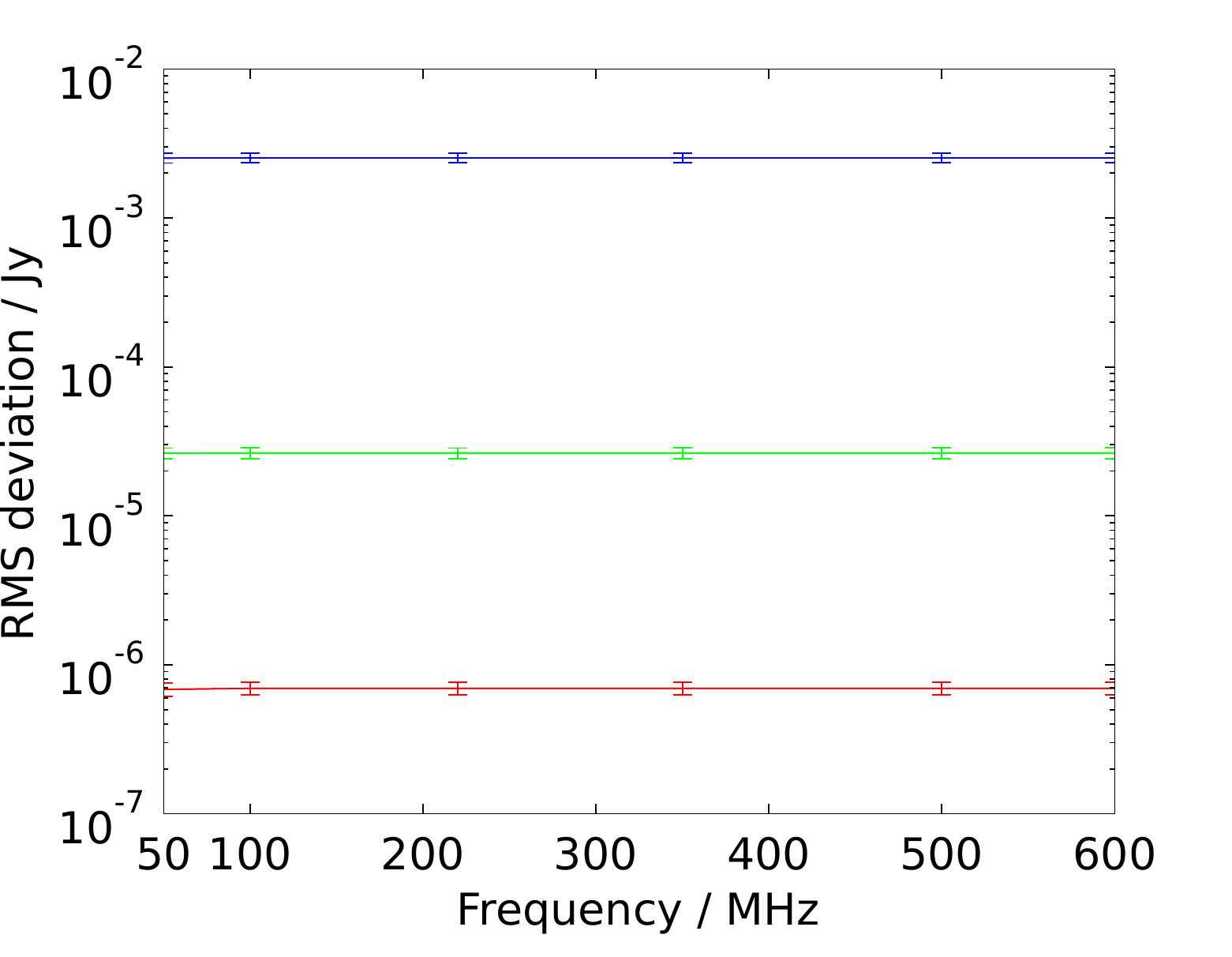}}
 \begin{center}
  \caption{The RMS deviation between images taken by an ideal telescope and one with errors for a sky model of a 3\,x\,3 grid of 1\,Jy point sources (model 1) in the station beam plotted against observation frequency. The separation between the sources is decreased with frequency, such that the source centres occupy the same pixels at every frequency. An example image from the ideal telescope is shown in Fig \ref{fig:ideal_dense_image}. \subref{fig:dense_amperr} Gain errors of 100\% are shown in blue, 10\% in green and 1\% in red. \subref{fig:dense_phaseerr} Phase errors of 10$^{\circ}$ are in blue (error bars $\times 100$) 1$^{\circ}$ in green (error bars $\times 20$) and 0.1$^{\circ}$ in red (error bars $\times 2$).}
  \end{center}
  \label{fig:dense}
\end{figure*}

The results for the dense grid sky model (model 1) are shown in Figs \ref{fig:dense_amperr} and \ref{fig:dense_phaseerr} are useful test case. The RMS deviation between the `ideal' and `error' images remains very nearly constant across the entire frequency range. As this sky model only contained sources in the station beam, where the source separation scaled perfectly with the field of view, this result is as expected; sparse stations cause differences in station beam sidelobes, rather than the main lobe of the beam. 

\begin{figure*}
\subfigure [~Simulated image of a dense grid of sources positioned within the station beam of the telescope. The sources are separated by 0.7$^{\circ}$ and the observation simulated at 50\,MHz. There are no phase or gain error effects included in this image.]{\label{fig:ideal_dense_image}\includegraphics[width=0.39\linewidth]{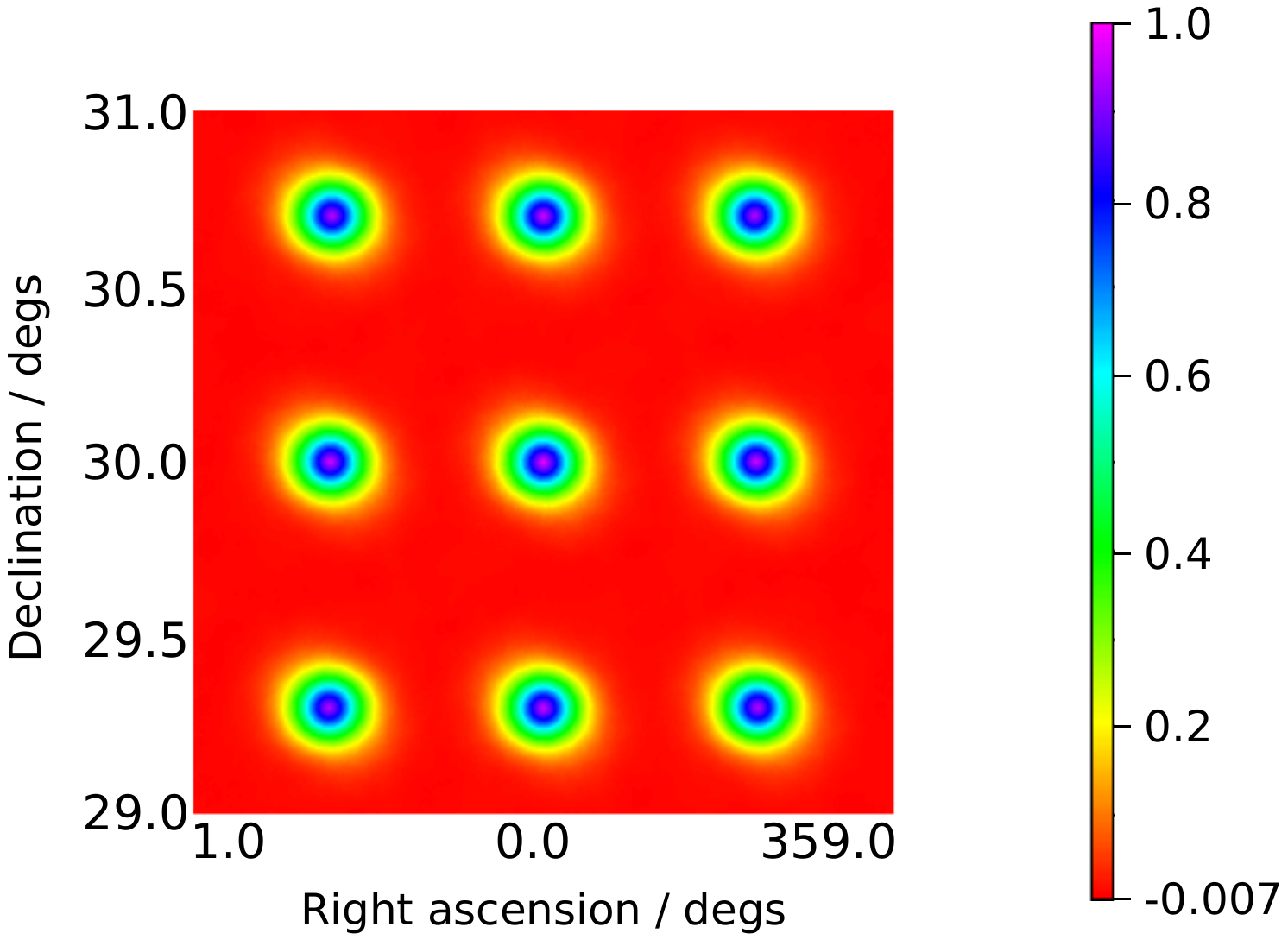}}
\subfigure [~The difference between image \ref{fig:ideal_dense_image} and an image taken with a telescope with element gain errors of 100\%.]{\label{fig:diff_dense_image}\includegraphics[width=0.39\linewidth]{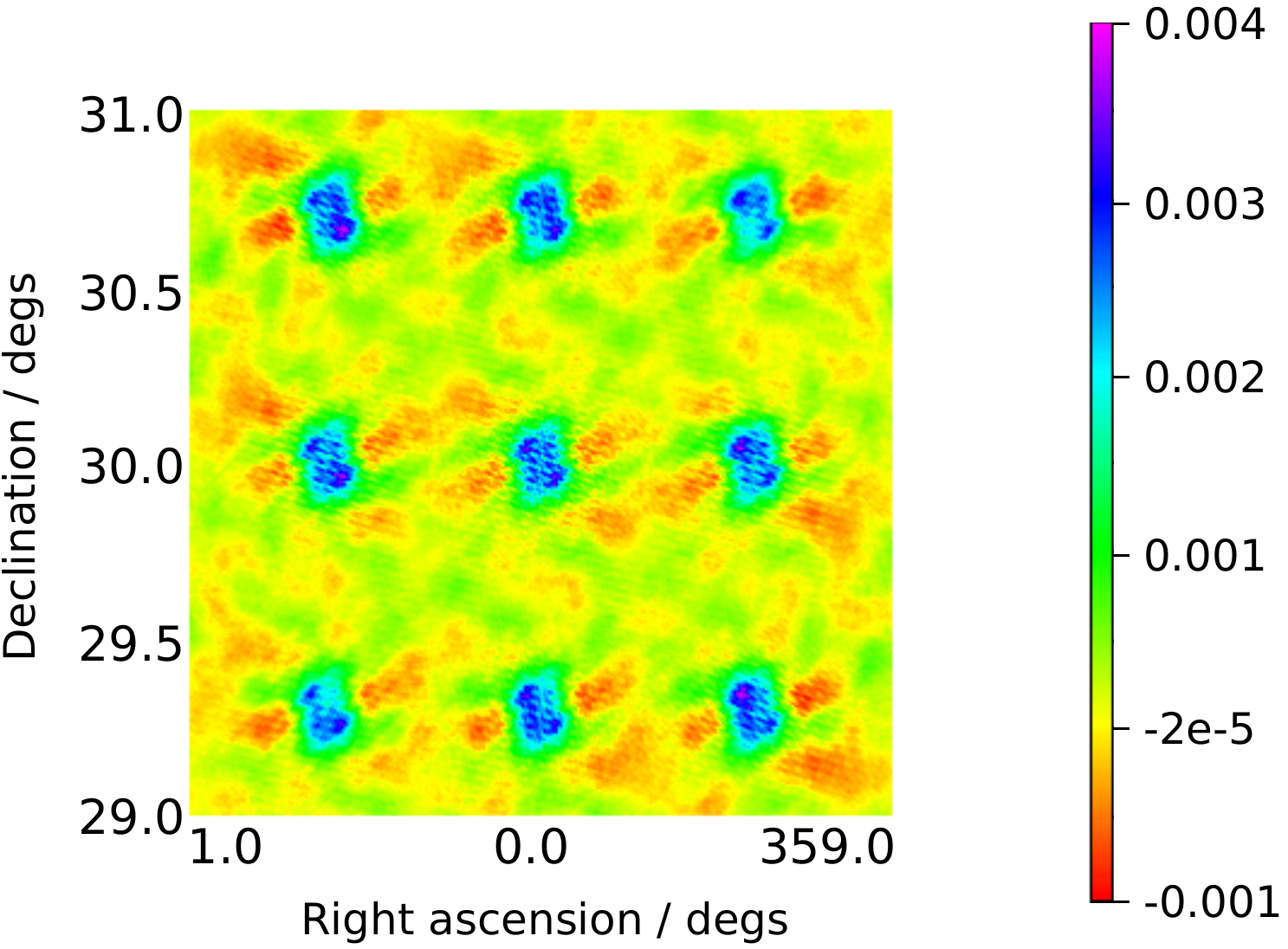}}

\subfigure [~Simulated image of the VLSS catalogue sky located in the sidelobes of the station beam pattern, with the phase centre pointed at an empty patch of sky. No sources are located within 2 x HPBW of the observation's phase centre. Image taken at 100\,MHz. There are no phase or gain error effects included in this image.]{\label{fig:hemi_ideal_image}\includegraphics[width=0.39\linewidth]{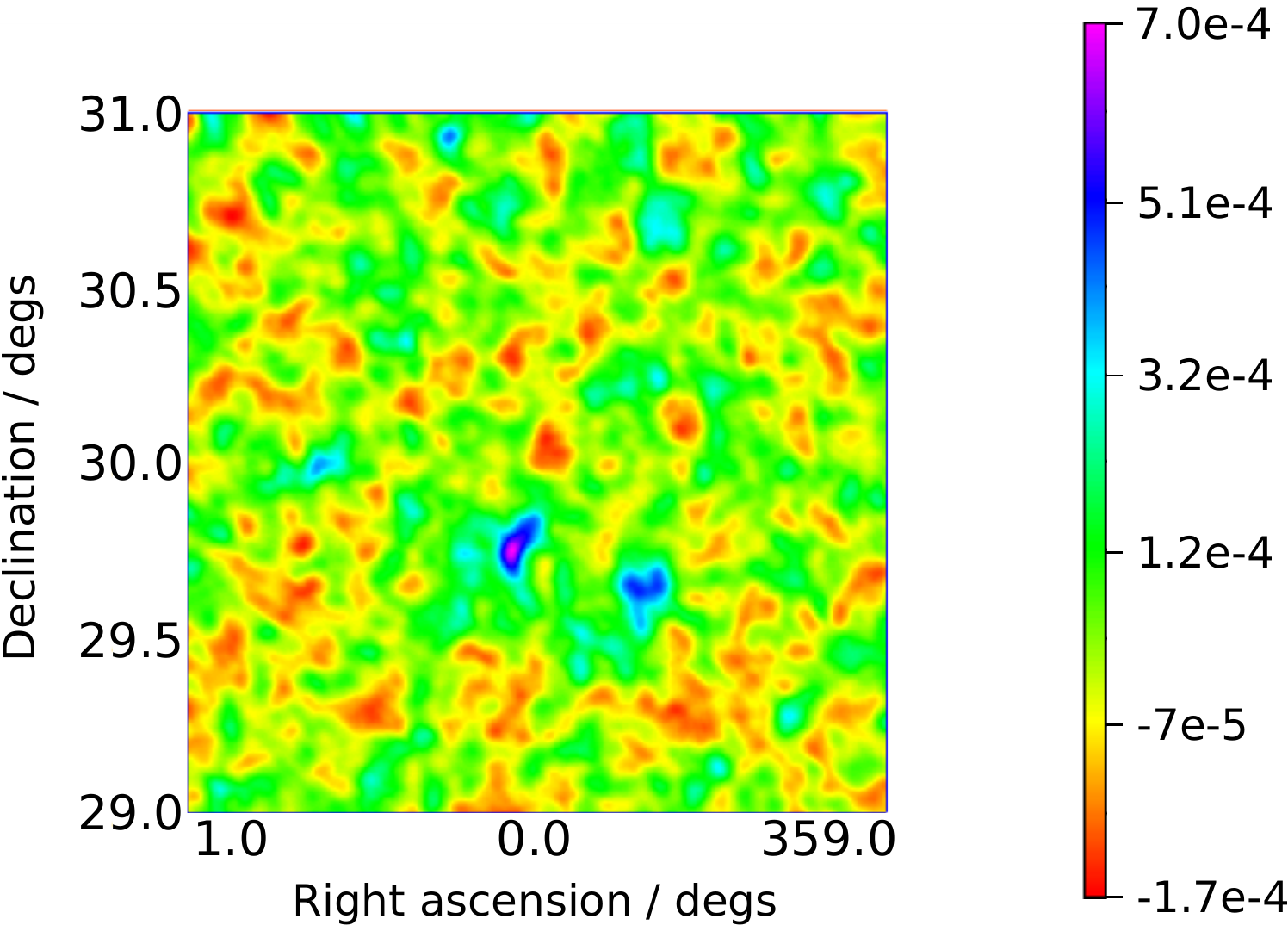}}
\subfigure [~The difference between image \ref{fig:hemi_ideal_image} and an image taken with a telescope with element gain errors of 100\%.]{\label{fig:hemi_diff_image}\includegraphics[width=0.39\linewidth]{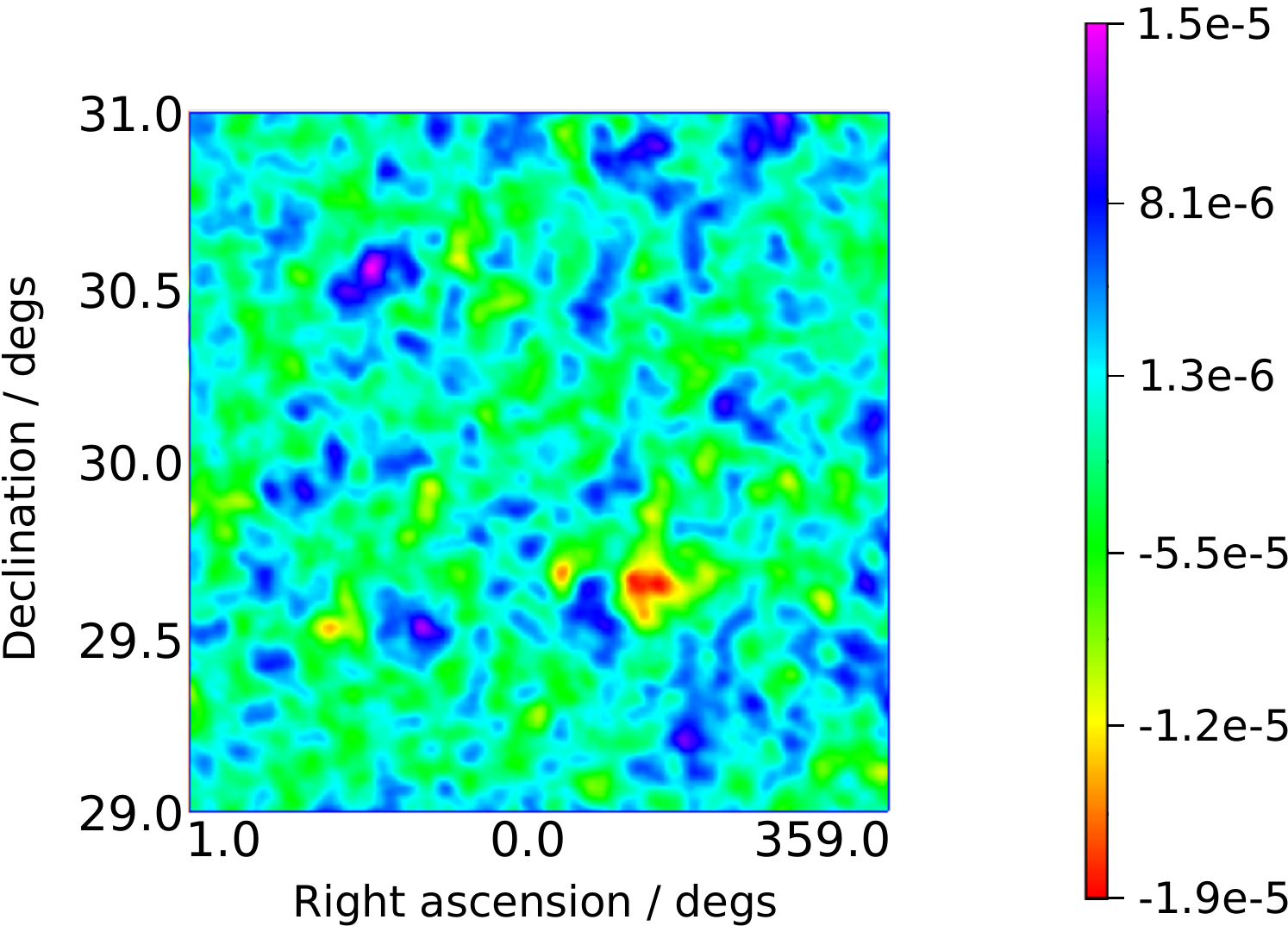}}
 \begin{center}
  \caption{\subref{fig:ideal_dense_image} and \subref{fig:hemi_ideal_image}: images from the simulated telescope. \subref{fig:diff_dense_image} and \subref{fig:hemi_diff_image}: the difference between images from a simulated ideal telescope with and without gain errors.}
  \end{center}
  \label{fig:sky_images}
\end{figure*}

An example of the difference between the ideal and error images for the dense sky model is shown in Fig \ref{fig:diff_dense_image}, alongside the ideal image  in Fig \ref{fig:ideal_dense_image}. The largest differences between the ideal and error images are found at the locations of the sources. This is not surprising, as these pixels measure the largest flux values; a $1\%$ error, for example, will cause a much greater absolute discrepancy here than in pixels recording near-zero flux levels. As the noise is not randomly distributed across the image, it may call into question the extent of the applicability of using RMS deviation as a metric, which should strictly be used for Gaussian noise. However, the RMS deviation is used here purely as a measure of the difference between the two images' respective pixel values; other metrics such as the total difference or maximum difference would show the same trend in results.

\begin{figure*}
\subfigure []{\label{fig:VLSS_beam_amperr}\includegraphics[width=0.43\linewidth]{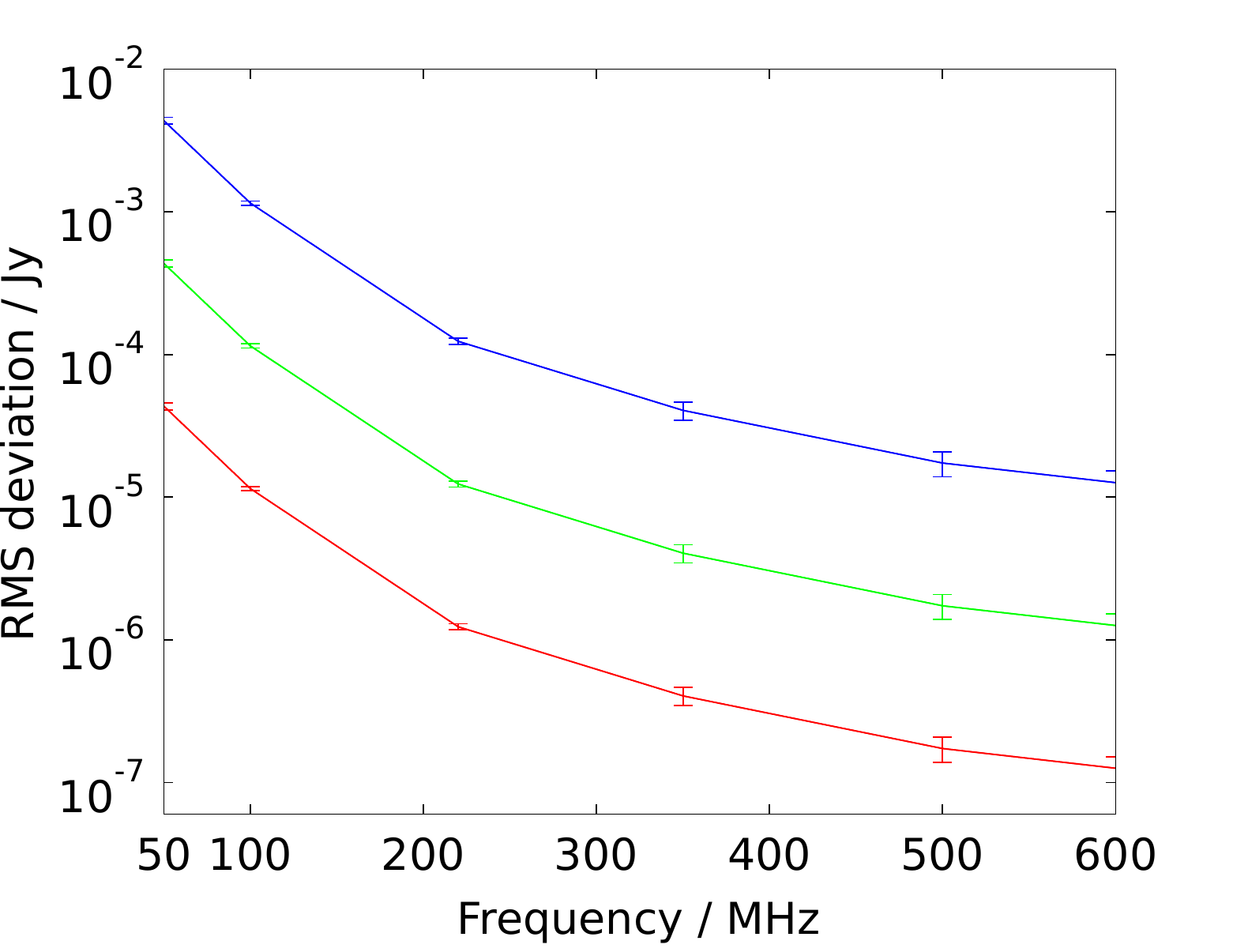}}
\subfigure []{\label{fig:VLSS_beam_phaseerr}\includegraphics[width=0.43\linewidth]{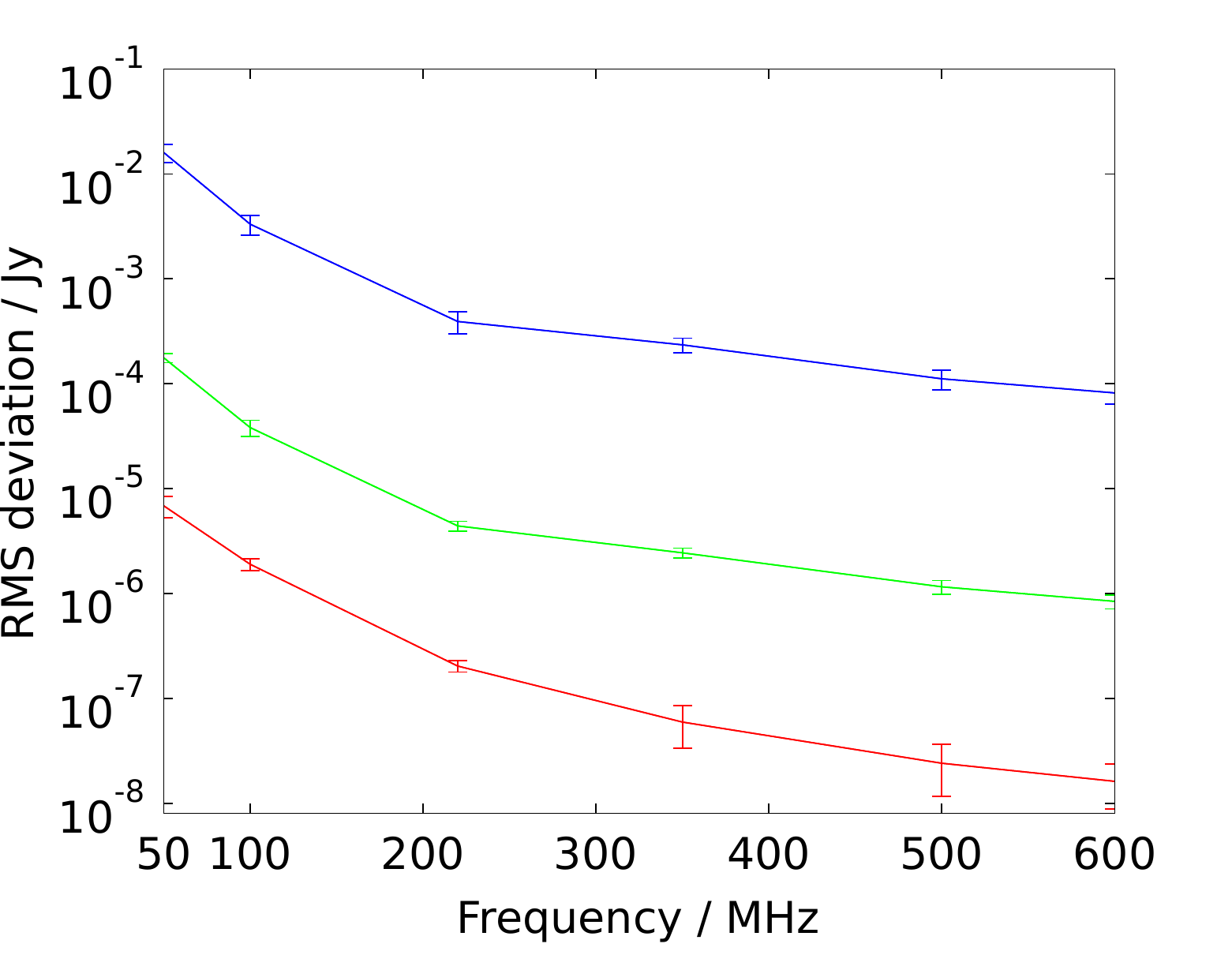}}
 \begin{center}
  \caption{Sky model from the VLSS catalogue; sources beyond the HPBW were removed from the sky model, as shown in Fig \ref{fig:VLSS_in_beam_FoV}. RMS deviation between images taken by an ideal telescope and one with errors. The simulated telescopes were pointed at the zenith. \subref{fig:VLSS_beam_amperr} Gain errors of 100\% are shown in blue, 10\% in green and 1\% in red. \subref{fig:VLSS_beam_phaseerr} Phase errors of 10$^{\circ}$ are in blue (error bars x 100), 1$^{\circ}$ in green (error bars x 20) and 0.1$^{\circ}$ in red (error bars x 10).}
  \end{center}
  \label{fig:VLSS_beam}
\end{figure*}

Figs \ref{fig:VLSS_beam_amperr} and \ref{fig:VLSS_beam_phaseerr} shows the results for sources in the station beam from the VLSS sky model (model 2). There is a trend of decreasing RMS deviation with increasing frequency. As the field of view decreases at higher frequencies, so do the number of sources in the field of view. As shown above, it is the pixels corresponding to the source locations that contribute the most to the RMS deviation, so fewer sources in the smaller beams leads to smaller RMS deviations. Furthermore, the spectral index of the VLSS sky model is set to -0.7, so the flux of the sources in the beam is diminished at the top of the frequency band relative to the flux at the bottom. As a semi-realistic sky model was used here, the RMS deviations give an approximation for the lower bound of source fluxes measurable from on-sky observations of 13.7\,s duration with 100\,kHz bandwidth.

\begin{figure*}
    \centering
    \includegraphics[width=0.7\textwidth]{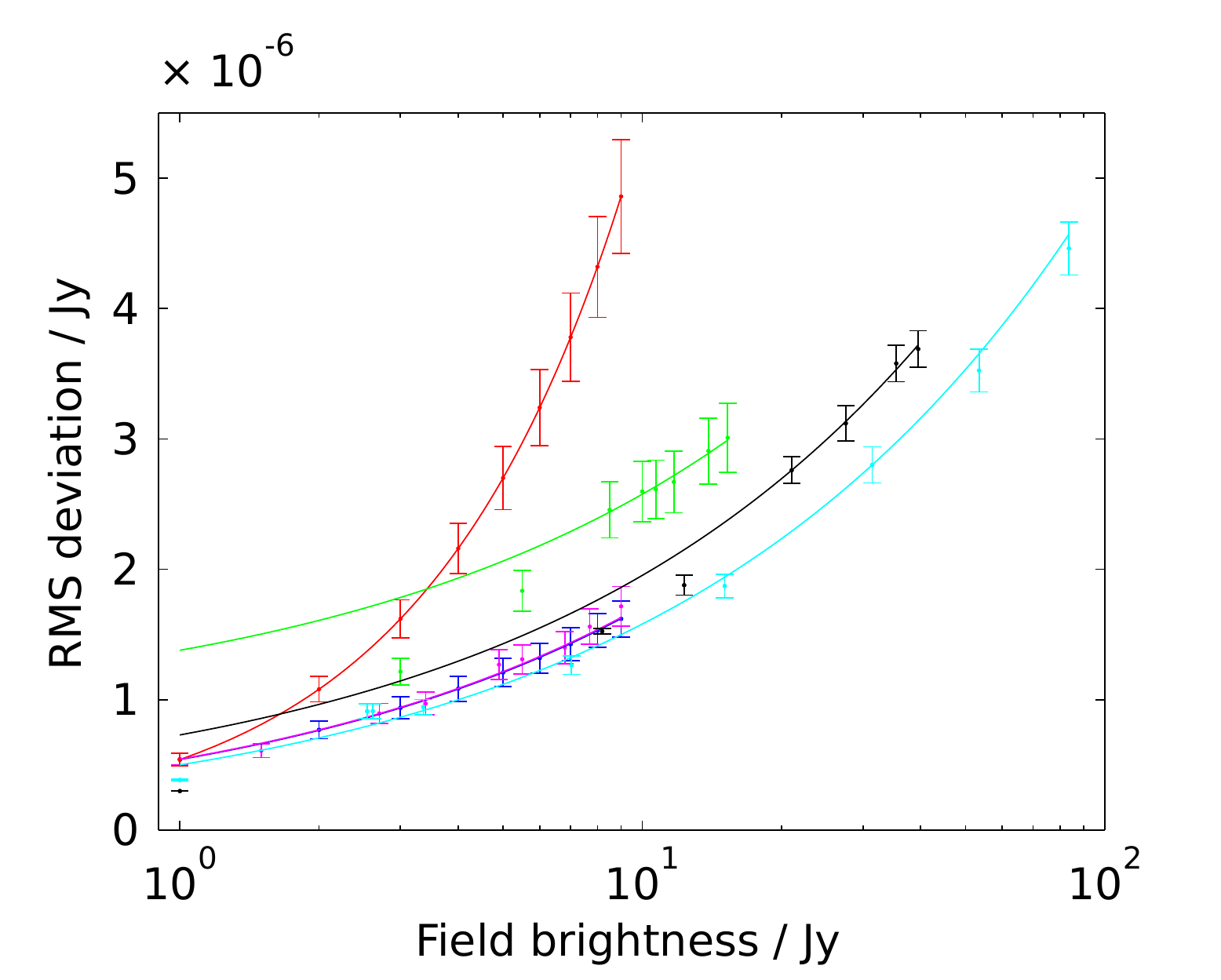}

    \caption{RMS deviation from residual in-field noise between ideal and error-telescope images, as a function of field brightness. The dark blue data points show the case of separate 1\,Jy sources positioned in a grid (as shown in Fig \ref{fig:ideal_dense_image}) in the HPBW of the beam (trend line $\propto \sqrt{B}$, where $B$ is the brightness in the field), whereas the red points are for the case when a single source of increasing brightness was used (trend line $\propto B$). The magenta and green data points use the same source locations as the blue data, but with varying source fluxes; the green data shows stronger fluxes and magenta weaker. The magenta data points follow a $\sqrt{B}$ scaling and, after an initial trend lying between linear and $\propto \sqrt{B}$, the green data points scale with $\sqrt{B}$, as highlighted by the fitted trend line.The light blue data shows a $\sqrt{B}$ relationship for a randomly generated source fluxes and positions within 80\,\% of the HPBW. The black data points show the values using a small number of sources from the VLSS catalogue, which also scales with $\sqrt{B}$ once a sufficient number of sources were included. Gain errors of 10\% were used at 100\,MHz.}
    \label{fig:rmsvsfieldflux}
\end{figure*}

To investigate the relationship between the total flux density in the field and the RMS deviation, further simulations were run, the results of which are shown in Fig \ref{fig:rmsvsfieldflux}. These simulations restricted all sources to be within 80\,\% of the HPBW, to minimise the differences in sensitivity depending on the source positions in the beam. 

When separate 1\,Jy sources were added to the field of view (in a gridded pattern, like the sources shown in Fig \ref{fig:ideal_dense_image}), their RMS deviation contributions added in quadrature, as the deviations were incoherent, resulting in a trend of RMS deviation $\propto \sqrt{B}$, where $B$ is the total field brightness (dark blue line, Fig \ref{fig:rmsvsfieldflux}). On the other hand, when the field brightness was increased by means of a single source with increasing flux, the RMS deviations were coherent and a linear relationship between RMS deviations and field brightness resulted (red line Fig \ref{fig:rmsvsfieldflux}).

Further simulations used the same source positions as the 1\,Jy gridded case (model 1), but with sources of differing flux levels. With weak sources (magneta line Fig \ref{fig:rmsvsfieldflux}), the $\sqrt{B}$ trend was followed, but with stronger fluxes (green data Fig \ref{fig:rmsvsfieldflux}), the relationship initially started as somewhere between $\propto \sqrt{B}$ and $B$, before settling into the $\propto \sqrt{B}$ trend when there were a greater number of sources. Simulations with the VLSS catalogue are also shown (black data Fig \ref{fig:rmsvsfieldflux}), the results of which generally follow the $\sqrt{B}$ trend, but when strong sources are added to the field, the increase is greater than $\propto \sqrt{B}$. Simulations using the VLSS catalogue are also shown in Fig \ref{fig:rmsvsfieldflux}, the results of which follow the $\sqrt{B}$ trend once a sufficient number of sources have been included in the field; the low field flux data points are more susceptible to be skewed by individual strong sources.

From these simulations it is possible to predict that the approximate RMS deviation produced by a large number of weak sources is given by
\begin{equation}
\Delta S = \left[ \sum_{i}^{\textrm{sources}} (k S_{i})^{2}\right]^{1/2}\,,
\label{equation:rmsvsfieldflux}
\end{equation}
where $S_{i}$ are the flux values for sources in the field and $k$ is a constant of proportionality. For these simulations, with a gain error of 10\%, $k = (5.4 \pm 0.1) \times 10^{-7}$, as determined by $\chi^{2}$ minimisation to the data sets (discounting first two and three data points for the strong souces in a grid data  (green line) and VLSS catalogue (black line) data sets respectively, where strong sources skew the trend away from $\sqrt{B}$). The accuracy of Equation (4) decreases when there are few sources, and when sources are far from the centre of the beam.

\subsection{Residual sidelobe noise}
\begin{figure*}
\subfigure []{\label{fig:grid_hemi_amperr}\includegraphics[width=0.43\linewidth]{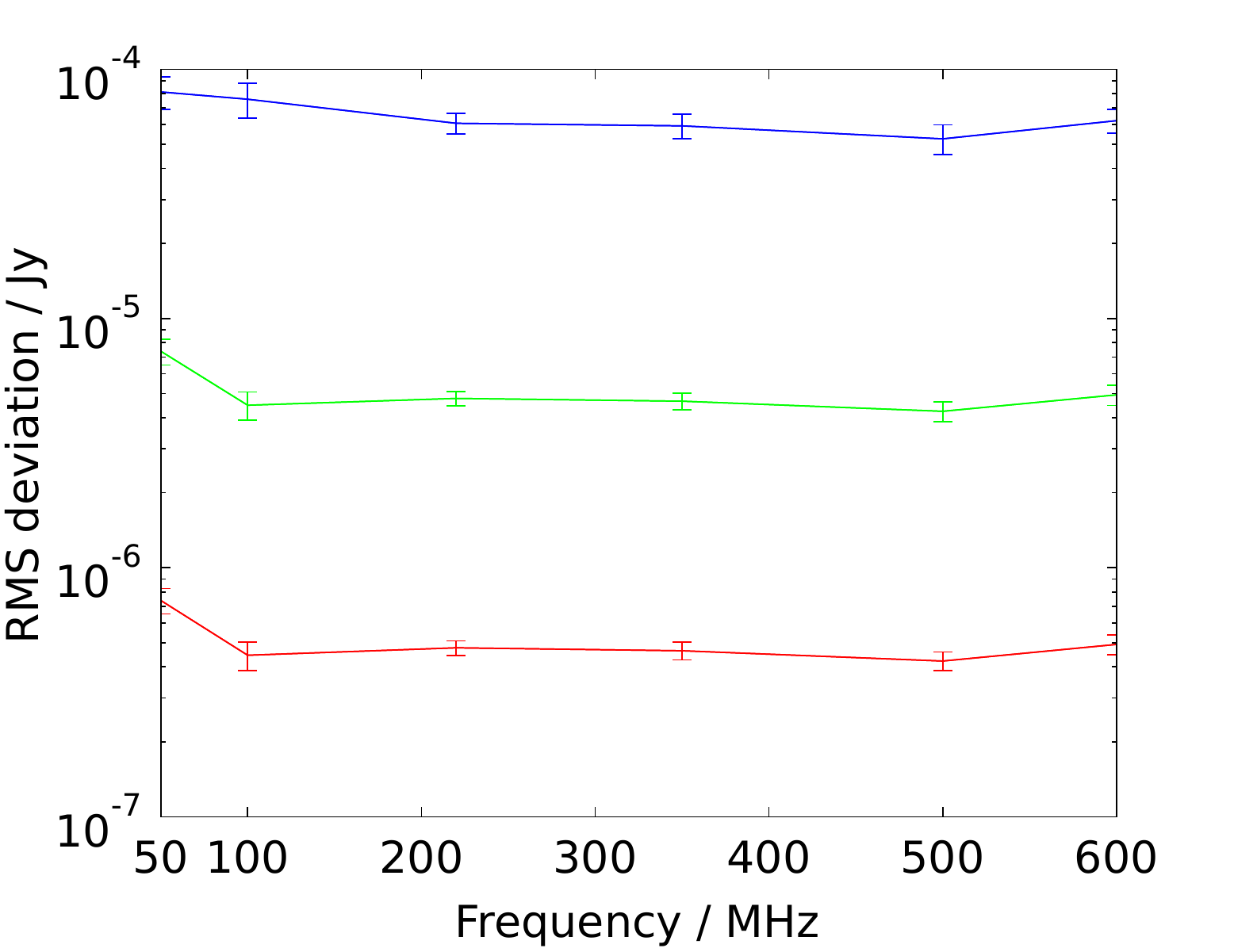}}
\subfigure []{\label{fig:grid_hemi_phaseerr}\includegraphics[width=0.43\linewidth]{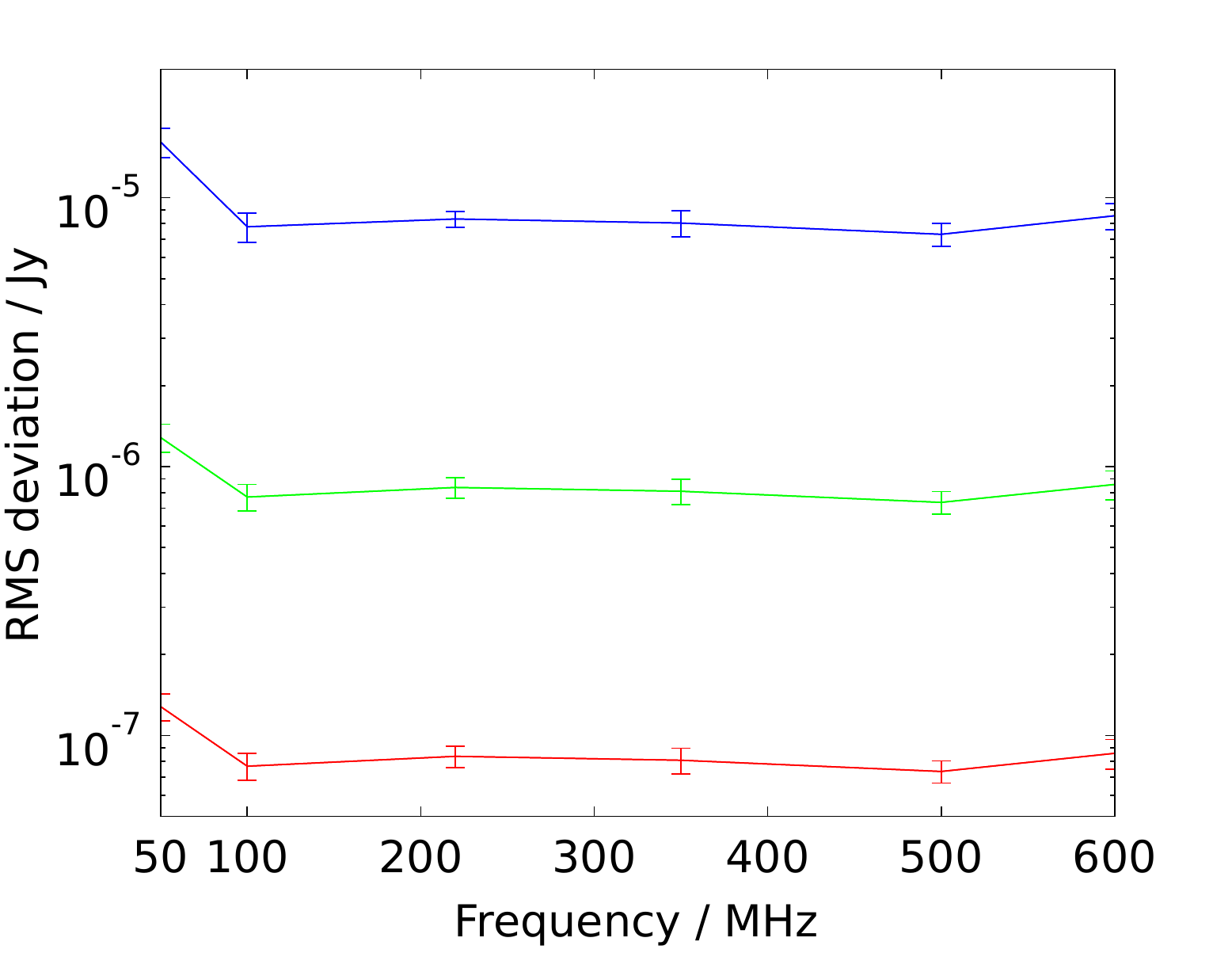}}
 \begin{center}
  \caption{Sky model: sparse grid of point sources in sidelobes. RMS deviation between images taken by an ideal telescope and one with errors. The simulated telescopes were pointed at an empty patch of sky with a sparse grid of 1\,Jy point sources, spread across the sky at 10$^{\circ}$ separations, in the sidelobes. An example image from the ideal telescope is shown in Fig \ref{fig:hemi_ideal_image}. \subref{fig:grid_hemi_amperr} Gain errors of 100\% are shown in blue, 10\% in green and 1\% in red. \subref{fig:grid_hemi_phaseerr} Phase errors. Phase errors of 10$^{\circ}$ are in blue, 1$^{\circ}$ in green and 0.1$^{\circ}$ in red.}
  \end{center}
  \label{fig:grid_hemi}
\end{figure*}

The other class of sky models had no sources within twice the HPBW of the station beam, but populated the rest of the sky with sources. Fig \ref{fig:hemi_diff_image} shows an example image of the difference between the ideal and error-case images, alongside the corresponding ideal image \ref{fig:hemi_ideal_image}. Figs \ref{fig:grid_hemi_amperr} and \ref{fig:grid_hemi_phaseerr} show the results for the gridded model (model 3), where sources were spaced at 10$^{\circ}$ intervals. The RMS deviation is seen to stay approximately constant across the frequency range, with the values slightly increasing and decreasing across the frequency range. As the pattern of the sidelobes varies with frequency, different sources move in and out of them, potentially explaining this trend. 

\begin{figure*}
\subfigure []{\label{fig:VLSS_hemi_amperr}\includegraphics[width=0.43\linewidth]{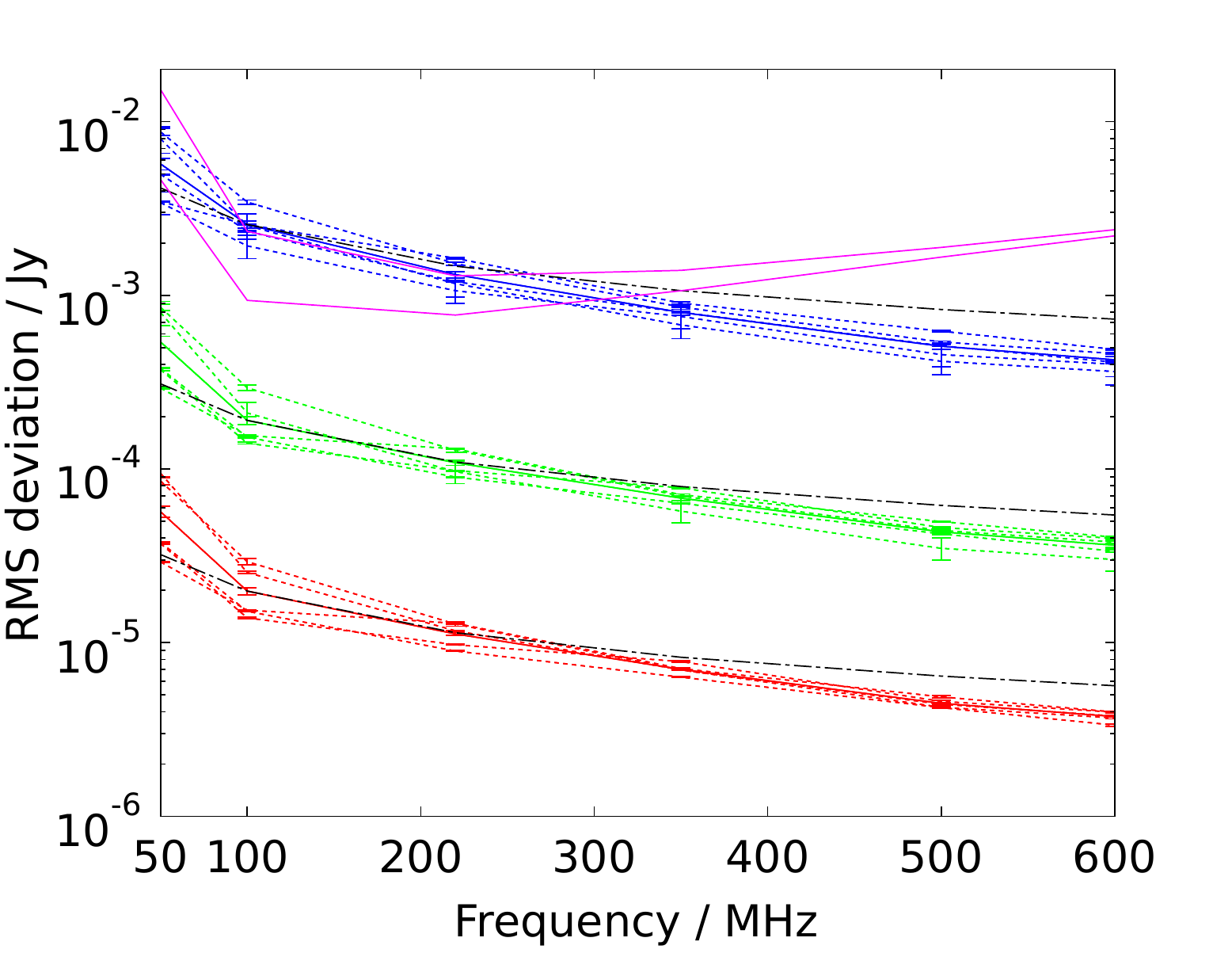}}
\subfigure []{\label{fig:VLSS_hemi_phaseerr}\includegraphics[width=0.43\linewidth]{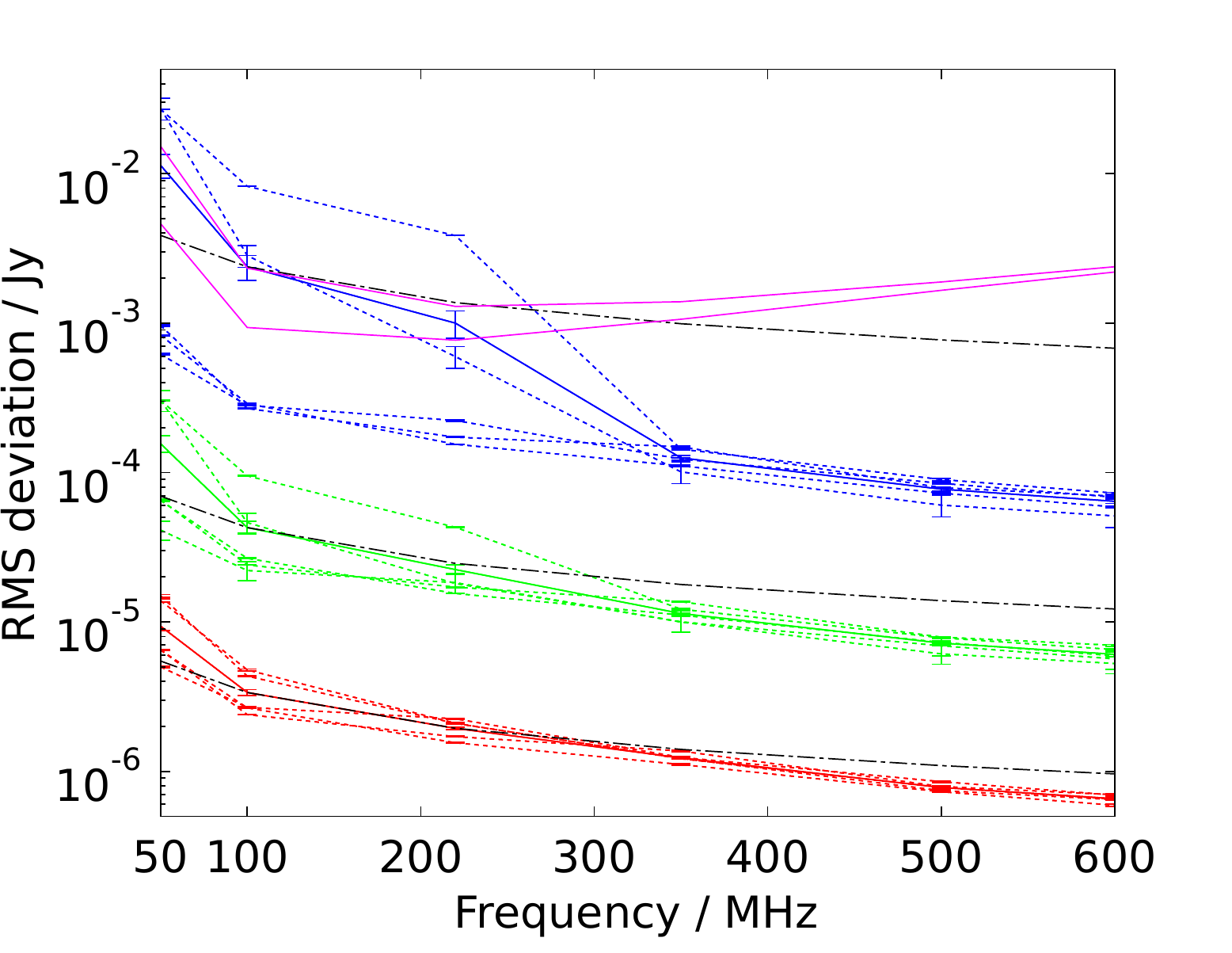}}
 \begin{center}
  \caption{RMS deviation between images taken by an ideal telescope and one with errors. The simulated telescopes were pointed at empty patches of sky with the VLSS point source sky model in the sidelobes. The size of the empty patch of sky was $2\,\times$\,HPBW of the station beam. Observations were taken in five different directions: (0,\,0), (10,\,10), (20,\,-15), (-10,\,0), (-5,\,-10), where the values are the (RA,\,Dec) away from the zenith in degrees. Different observing directions are shown by the dashed lines, and the mean is shown by the solid lines. The black dot-dashed lines show the RMS values if they were to decrease according to the spectral index (-0.7) of the sky model, referenced to the mean RMS value for all the observing directions at 100\,MHz. The magenta lines shows the probable range of thermal noise level at each frequency. \subref{fig:VLSS_hemi_amperr} Gain errors. Gain errors of 100\% are shown in blue, 10\% in green and 1\% in red. \subref{fig:VLSS_hemi_phaseerr} Phase errors. Phase errors of 10$^{\circ}$ are in blue, 1$^{\circ}$ in green and 0.1$^{\circ}$ in red.} 
  \end{center}
  \label{fig:VLSS_hemi}
\end{figure*}

The other RSN simulations used the VLSS point-source model (model 4). As the exact number of sources and their position with respect to the sidelobes was judged to affect the measured RMS deviations, observations were made in five different directions: (0,\,0), (10,\,10), (20,\,15), (-10,\,30) and (-5,\,20) (values quoted are degrees in RA and Dec away from the zenith). As shown in Fig \ref{fig:VLSS_hemi_amperr} and \ref{fig:VLSS_hemi_phaseerr}, a spread of results can be seen depending on the beam direction, but each follows the same trend of decreasing RMS deviation with frequency. The RMS deviations decrease approximately by a factor of 10 across the frequency range, with around 50\,\% of the drop occurring between 50 and 100\,MHz. This is partially attributable to the decreased sky brightness at higher frequencies, as shown by the black dot-dashed lines in Figs \ref{fig:VLSS_hemi_amperr} and \ref{fig:VLSS_hemi_phaseerr}. These lines display the expected RMS values according to the power law spectral index of -0.7, relative to the RMS deviation at 100\,MHz. The rest of the reduction with frequency could be explained by the reduced average sidelobe level across the sky at higher frequencies; high-order sidelobes tend to be lower in sensitivity than near-in sidelobes, and the higher the observing frequency, the greater the number of sidelobes an array will have. Whilst grating lobes will disappear below 110\,MHz, the critical sampling frequency, the near-in sidelobe will compensate for this by covering large solid angles with a sensitivity that is a significant fraction of that of the main beam.

The magenta lines in Figs \ref{fig:VLSS_hemi_amperr} and \ref{fig:VLSS_hemi_phaseerr} show the approximate range of the thermal noise floor of the telescope model for a 13.7\,s integration with 100\,kHz bandwidth, as calculated by Equations (1) to (3). Measurements were taken of both typical and relatively cold patches of sky. The thermal noise limit shows that gain errors of 10\% and phase errors of 1$^{\circ}$ are comfortably within the thermal noise level, but gain errors of 100\% and phase errors of 10$^{\circ}$ can produce noise levels exceeding the thermal noise. From inspection of these results, an approximate upper bound of ${\sim}\,0.2$ in gain error and ${\sim}\,5^{\circ}$ in phase error would keep the RMS deviations below the thermal noise levels at all frequencies for snapshot images. It is important to stress that this is a best-case scenario, assuming the position and flux of all sources are known perfectly at each observation frequency.

The RMS deviations in both Figs \ref{fig:VLSS_beam_amperr}, \ref{fig:VLSS_beam_phaseerr}, \ref{fig:VLSS_hemi_amperr} and \ref{fig:VLSS_hemi_phaseerr} indicate that it is the low frequencies at which SKA1-low should be most concerned to ensure the telescope can meet the sensitivities required by the science goals. It seems that at low frequencies, the large beam sizes, and hence large number of sources in beams and the prominent sidelobes, result in greater distortions of the visibilities due to errors in the telescope model. Based on these results, if images can be satisfactorily produced at low frequencies, then the higher frequencies should not present greater difficulties.

\begin{figure*}
    \centering
    \includegraphics[width=0.7\textwidth]{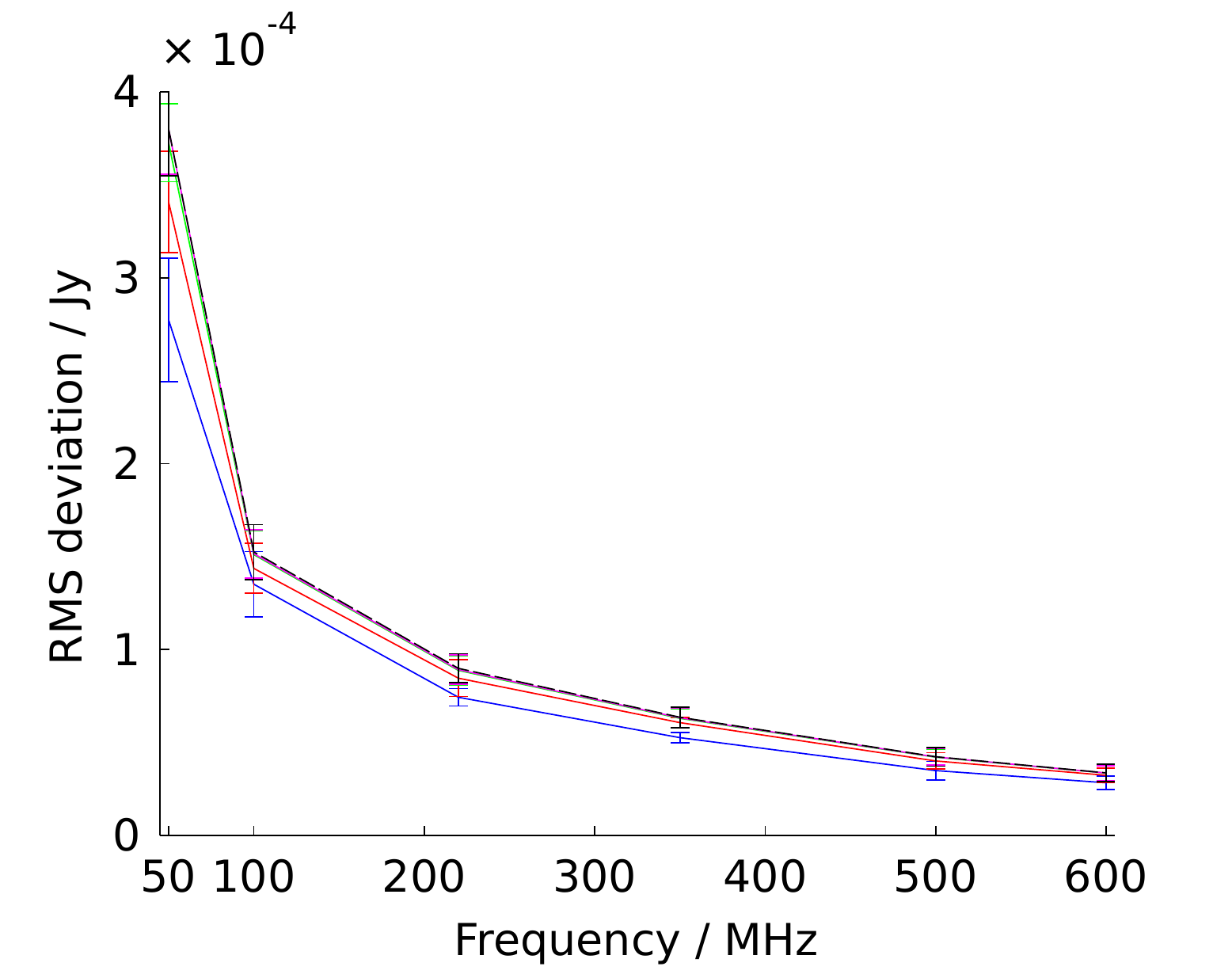}
    \caption{RMS deviation for sky models with different minimum fluxes in the sky model. The VLSS sky model was used, but only the brightest 1\,\% (i.e. $\textgreater14$\,Jy at 70\,MHz), 10\,\% ($\textgreater 3.7$\,Jy), 50\,\% ($\textgreater 1.2$\,Jy) and 100\,\% of sources were used. These are shown by the blue, red, green and black data points respectively; the magenta data points shows the RMS when all sources within double the weakest source flux in the VLSS sky model (i.e. sources $\textless 0.79$\,Jy) were removed (there is only a small difference between the 50\,\%, 100\,\% and double the weakest source data). A gain error of 10\% was used.}
    \label{fig:cropsky}
\end{figure*}

To assess the impact of the strongest sources on the RMS deviation, simulations were run which removed varying fractions of the weakest sources in the sky model. Fig \ref{fig:cropsky} shows that the RMS deviations are dominated by the brightest 10\,\% of sources in the sky. This is encouraging, as in reality no sky model will be able to perfectly characterise all the sources in the sky.

\subsection{Effect of integration time}

\begin{figure*}[t!]
    \centering
    \subfigure []{\includegraphics[width=0.7\linewidth]{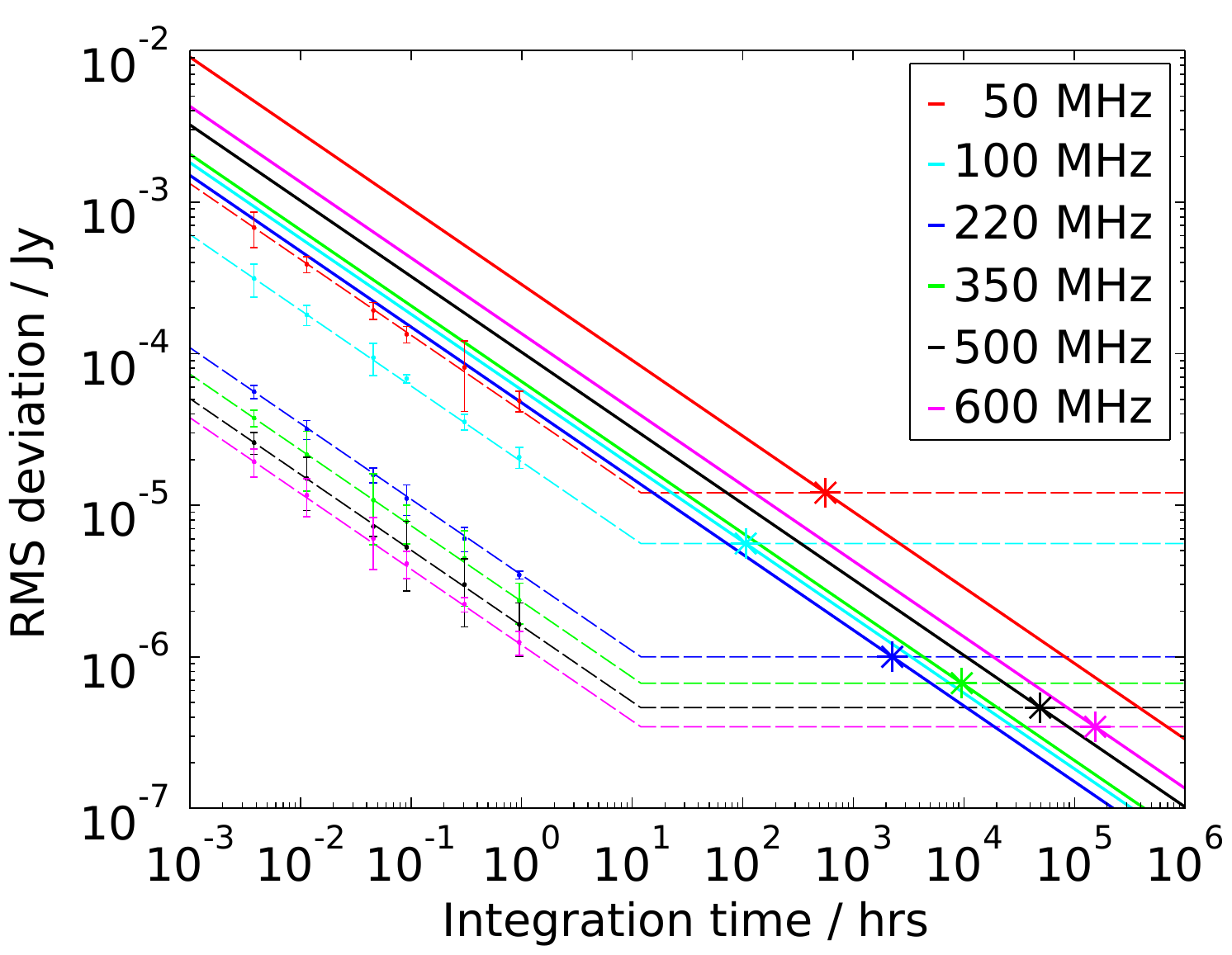}}
    \vspace{2em}
        \subfigure []{\includegraphics[width=0.45\linewidth]{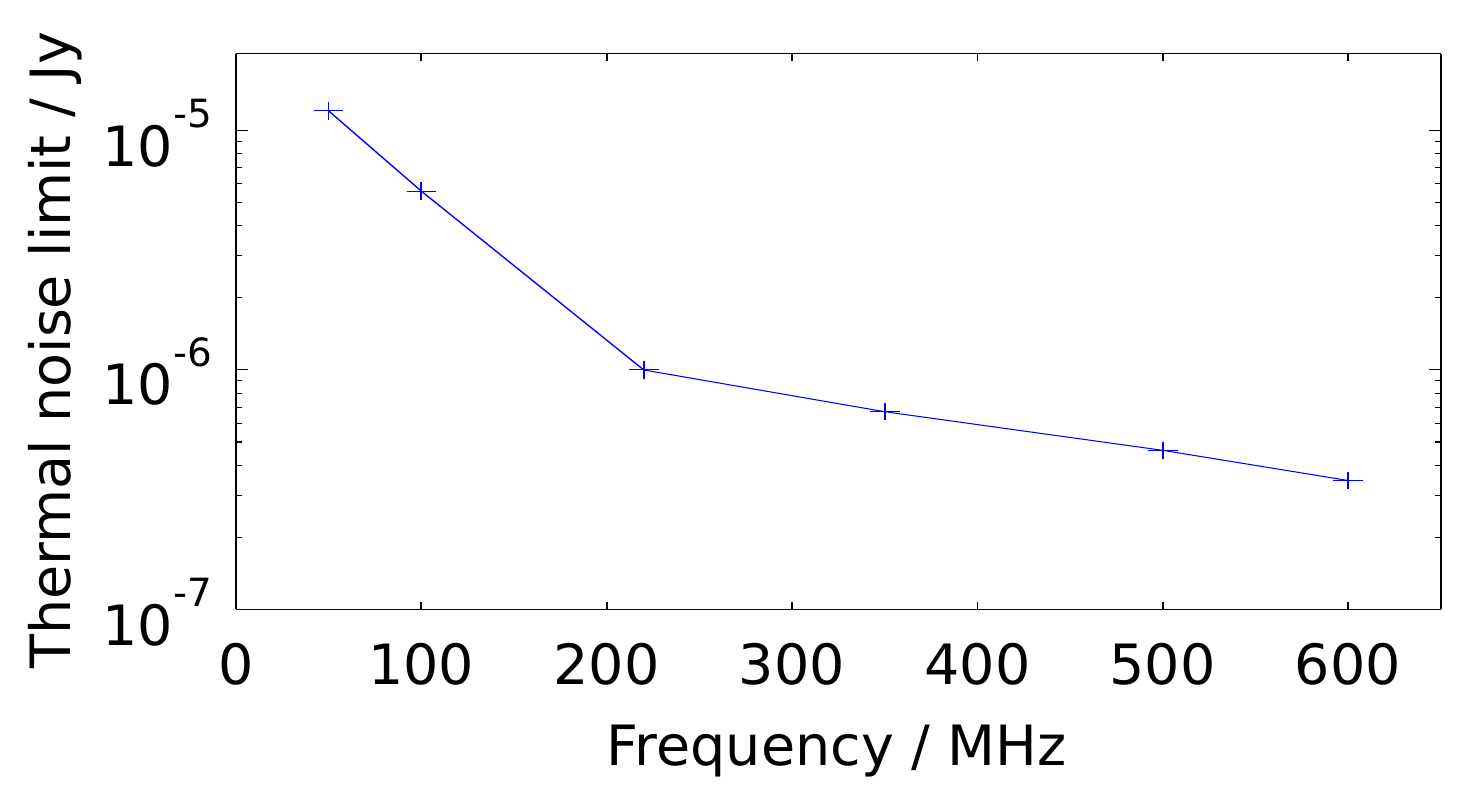}}
         \subfigure []{\includegraphics[width=0.45\linewidth]{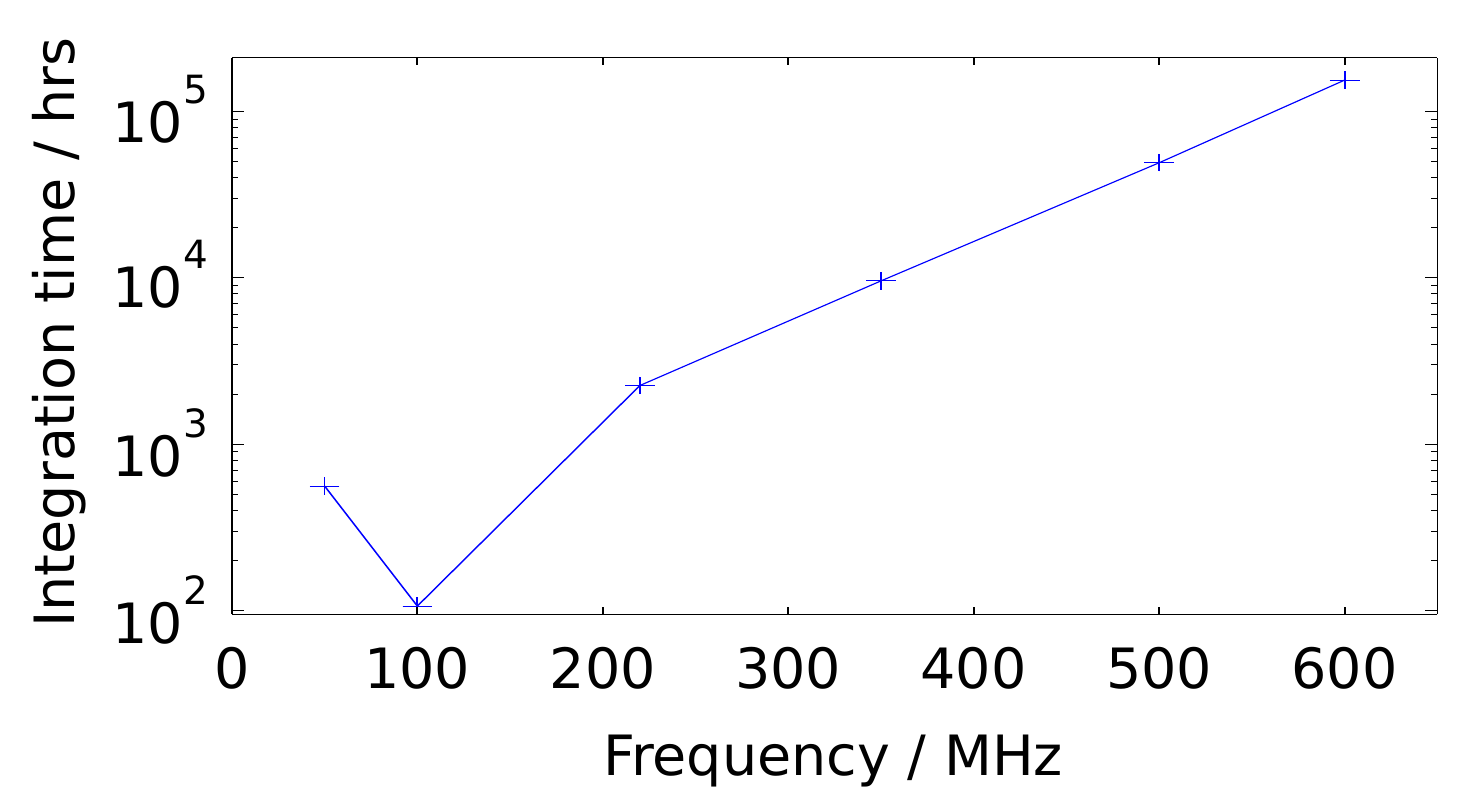}}
    \caption{(a) RMS deviation between the ideal and error image as a function of integration time shown by the data points (error bars x10) for a range of frequencies. The dashed lines through the data points show a $t^{-0.5}$ trend line. Thermal noise levels are shown in solid lines. The stars highlight the integration time at which the thermal noise drops beneath the RMS deviation; these values are also plotted as a function of frequency in (b) for clarity. Different coloured lines show the RMS deviation and thermal noise at different frequencies from $50 \-- 600$ MHz. Simulations were run with the VLSS sky model with an empty area of sky in the main beam. A gain error of 10\% was used; equivalent results are obtained for a phase errors of ${\sim}\,6^{\circ}$. (c) shows the integration time required to reach the thermal noise limit as a function of frequency, using data extracted from (a). }
    \label{fig:timesteps}
\end{figure*}

A further simulation was carried out to test the effect of integration time on the RMS deviations. Fig \ref{fig:timesteps} shows how the RMS deviation level changed for extended integration times, of up to 57\,mins, across the frequency band. The results follow the expected $t^{-0.5}$ progression, where $t$ is the observation duration, across the frequency range. The 50\,MHz results suggest a possible deviation from the $t^{-0.5}$ trend, reducing the improvement in noise with time, but longer integration time simulations would be required to establish this.

The dashed lines in Fig \ref{fig:timesteps} extrapolate the RMS deviation to 12 hours integration according to the $t^{-0.5}$ reduction. Beyond 12 hours, the \textit{uv} coverage no longer increases as the earth rotates, rather it repeats itself, which diminishes the benefit from extended observations. The thermal noise level, shown in solid lines, also decreases by $t^{-0.5}$ as it is simply Gaussian noise, but without the twelve hour restriction. Once the thermal noise decreases below the RMS deviation it limits the usefulness of longer integrations. 

By interpolating the impact of gain and phase errors on the RMS deviation in Figs \ref{fig:VLSS_hemi_amperr} and \ref{fig:VLSS_hemi_phaseerr}, the RMS deviation created by a gain error of 10\% is equivalent to a phase error of ${\sim}\,6^{\circ}$. Hence, from Fig \ref{fig:timesteps} it can be stated that for gain errors of 10\%, or phase errors of ${\sim}\,6^{\circ}$, the achievable thermal noise limit is between ${\sim}\,10-0.3$\,\textmu Jy and reached after ${\sim}\,100-100\,000$\,hrs, depending on frequency. The shortest achievable integration time is for 100\,MHz, with a thermal noise limit of 6\,\textmu Jy achieved after 100\,hr integration. At 50\,MHz the best noise level is higher, at 10\,\textmu Jy, but reached after a much longer integration time (${\sim}\,600$\,hr) due to the much higher system temperature at 50\,MHz. Above 100\,MHz the best achievable noise level drops rapidly.  Improving the gain and phase errors would further decrease the thermal noise limit.

An important caveat is that 57\,mins is a limited time period compared to the multi-hour observations often performed with interferometers. Longer duration simulations were not practical owing to available compute resources, but it would be desirable to simulate much greater integration times to verify the accuracy of the extrapolations.

As with all simulations, these have their limitations. The performance of electronic components are not explicitly modelled, rather the noise introduced by components such as the LNAs (low noise amplifiers) is assumed to be included in the gain and phase errors introduced on an antenna basis. This neglects the fact that electronics noise is likely to vary with frequency, especially across a wide bandwidth of 50 \-- 650\,MHz.

Other errors in the station beams are also not explicitly modelled, with these simulations assuming that no extra uncertainties are added when the antenna signals are combined to create station level beams. Quantisation and noise from electronic components will prevent this idealised scenario.

It is also worth noting that for time-dependent errors, such as imprecise antenna locations, the resultant phase errors will increase with frequency. However, it is difficult to know precisely the contribution of time-dependent and time-independent phase errors in the final signal, so modelling this effect is not necessarily straightforward. These simulations assume the phase errors specified are the same across the frequency range.

The VLSS sky model used for these simulations is an incomplete model of the sky. The VLSS model is a point-source sky model, hence extended sources, including the galactic plane, were not accounted for within the simulations. It is not clear quite how these would have affected the results. Furthermore, applying a spectral index of -0.7 to model the change in flux with frequency is a simplification; in reality source fluxes will vary in a range of ways with frequency. Finally, the differences in RMS of the beam directions in Fig \ref{fig:VLSS_hemi_phaseerr} suggests that very bright sources can produce RMS differences that are an order of magnitude bigger than what we see here. As stated previously, the six brightest sources were removed perfectly from the VLSS catalogue used as a sky model; the RSN could potentially rise by amounts greater than an order of magnitude if one of these sources were located in a strong sidelobe.

\section{Conclusions}
These simulations suggest that if the signal processing is of a high enough accuracy so that errors are dominated by antenna gain and phase errors, the effect of uncorrelated errors in the beam pattern of SKA1-low will make imaging more difficult at the low frequency end of the band compared to the high frequency. This can be attributed to larger field of view and increased flux from sources at low frequencies. 

The RMS deviations between ideal and error images decrease with increasing frequency and increasing magnitude of the errors. These simulations indicate that the thermal noise limit on images will range between ${\sim}\,10-0.3$\,\textmu Jy and reached after ${\sim}\,100-100\,000$\,hrs, for observations from 50\--600\,MHz.

The results also suggest that, in order to mitigate the effect of large beam areas on imaging at the lowest SKA1-low frequencies, consideration be given to schemes which increase the effective station size at low frequency.

\subsubsection*{Acknowledgements}
The authors would like to thank Nima Razavi-Ghods for helpful discussions.

The work presented here made use of the EMERALD HPC facility provided by the e-Infrastructure South Centre for Innovation (EPSRC Grant ref EP/K000144/1, EP/K000136/1).

The authors would like to acknowledge the use of the Advanced Research Computing (ARC) in carrying out this work.

\bibliography{imaging_paper_bib}
\end{document}